\documentclass[sigconf]{aamas} 
\usepackage{balance} % for balancing columns on the final page

\usepackage[utf8]{inputenc} % allow utf-8 input
\usepackage[T1]{fontenc}    % use 8-bit T1 fonts
\usepackage{hyperref}       % hyperlinks
\usepackage{url}            % simple URL typesetting
\usepackage{booktabs}       % professional-quality tables
\usepackage{amsfonts}       % blackboard math symbols
\usepackage{nicefrac}       % compact symbols for 1/2, etc.
\usepackage{microtype}      % microtypography
\usepackage{amsthm}
\usepackage{xcolor}
\usepackage{bm}
\usepackage{mathrsfs}
\usepackage{algorithmic}
\usepackage{siunitx}
\usepackage{algorithm, float}
\usepackage{graphicx}
\usepackage{xcolor}
\usepackage{nccmath}
\usepackage{comment}
\usepackage{stackengine}
\usepackage{float}
\usepackage{subfig}

\setcopyright{ifaamas}
\acmConference[AAMAS '21]{Proc.\@ of the 20th International Conference on Autonomous Agents and Multiagent Systems (AAMAS 2021)}{May 3--7, 2021}{Online}{U.~Endriss, A.~Now\'{e}, F.~Dignum, A.~Lomuscio (eds.)}
\copyrightyear{2021}
\acmYear{2021}
\acmDOI{}
\acmPrice{}
\acmISBN{}

\acmSubmissionID{120}

\title[Partially Observable Mean Field Reinforcement Learning]{Partially Observable Mean Field Reinforcement Learning}

\author{Sriram Ganapathi Subramanian}
\affiliation{
  \institution{University of Waterloo}
  \city{Waterloo, Canada}}
\email{s2ganapa@uwaterloo.ca}

\author{Matthew E. Taylor}
\affiliation{
  \institution{University of Alberta, Dept.~of Computing Science\\Alberta Machine Intelligence Institute (Amii)}
  \city{Edmonton, Canada}
  }
\email{matthew.e.taylor@ualberta.ca}

\author{Mark Crowley}
\affiliation{
  \institution{University of Waterloo}
  \city{Waterloo, Canada}}
\email{mcrowley@uwaterloo.ca}

\author{Pascal Poupart}
\affiliation{
  \institution{University of Waterloo, Waterloo, Canada}
  \city{Vector Institute, Toronto, Canada}}
\email{ppoupart@uwaterloo.ca}

\begin{abstract}
Traditional multi-agent reinforcement learning algorithms are not scalable to environments with more than a few agents, since these algorithms are exponential in the number of agents. Recent research has introduced successful methods to scale multi-agent reinforcement learning algorithms to many agent scenarios using mean field theory.  Previous work in this field assumes that an agent has access to exact cumulative metrics regarding the mean field behaviour of the system, which it can then use to take its actions. In this paper, we relax this assumption and maintain a distribution to model the uncertainty regarding the mean field of the system. We consider two different settings for this problem. In the first setting, only agents in a fixed neighbourhood are visible, while in the second setting, the visibility of agents is determined at random based on distances. For each of these settings, we introduce a $Q$-learning based algorithm that can learn effectively. We prove that this $Q$-learning estimate stays very close to the Nash $Q$-value (under a common set of assumptions) for the first setting. We also empirically show our algorithms outperform multiple baselines in three different games in the MAgents framework, which supports large environments with many agents learning simultaneously to achieve possibly distinct goals. 

\end{abstract}


\newtheorem{theorem}{Theorem}
\newtheorem{lemm}{Lemma}
\newtheorem*{theorem*}{Theorem}
\newtheorem{theorem2}{Theorem}

\newtheorem{assumption}{Assumption}

\newcommand\E{\mathbb E}

\newenvironment{proofsk}{%
  \renewcommand{\proofname}{Proof (Sketch)}\proof}{\endproof}
\def\delequal{\mathrel{\ensurestackMath{\stackon[1pt]{=}{\scriptstyle\Delta}}}}

\begin{document}

%%% The following commands remove the headers in your paper. For final 
%%% papers, these will be inserted during the pagination process.

\pagestyle{fancy}
\fancyhead{}

%%% The next command prints the information defined in the preamble.

\maketitle 

%%%%%%%%%%%%%%%%%%%%%%%%%%%%%%%%%%%%%%%%%%%%%%%%%%%%%%%%%%%%%%%%%%%%%%%%

\section{Introduction}
Multi-agent systems involve several learning agents that are learning simultaneously in an environment to solve a task or satisfy an objective. These agents may have to compete or cooperate with each other in the given context. Multi-agent systems are non stationary \cite{hernandez2019survey}, making it hard to derive learning policies that are as effective as in the single agent context. As the number of learning agents increases, the possible number of learning situations in the environment increases exponentially. Many algorithms introduced in the Multi-Agent Reinforcement Learning (MARL) literature suffer from scalability issues \citep{foerster2018learning,lanctot2017unified}, and hence are typically not well suited for environments in which agents are infinite in the limit, called \emph{many agent systems}. Mean field theory has been used to scale MARL to many agent scenarios in previous research efforts \citep{yang2018learning,pmlr-v80-yang18d}, most of which have assumed some notion of aggregation that is made available by an engine or is directly observable in the environment. For example, Guo et al.\ \cite{guo2019learning} assume that a population distribution parameter can be obtained from the game engine and Yang et al.\ \cite{pmlr-v80-yang18d} assume that the mean action of all agents in the environment can be observed directly by all agents. 

Partial observability is an important research area in single agent reinforcement learning (RL) \citep{hausknecht2015deep,karkus2017qmdp,zhu2017improving}, but these advances are not applicable to the many agent RL paradigm, since the stationary environment assumption is broken. Also, partial observation in single agent RL corresponds only to partial observability of state features, but in multi-agent systems this could also correspond to partial observability of other agents.

This paper relaxes the assumption that agents observe the aggregate state variable in a mean field update. Instead, we maintain a belief over the aggregate parameter that is used to help agent action selection. We focus on discrete state and action space Markov decision processes (MDPs) and modify the update rules from Yang et al.\ \cite{pmlr-v80-yang18d} to relax the assumptions of (1) global state availability and (2)  exact mean action information for all agents. We consider two settings in this paper. 
The \textit{Fixed Observation Radius (FOR)} setting assumes that all agents in each agent's small field of view are always observed (and those outside the radius are not).
The \textit{Probabilistic Distance-based Observability (PDO)} setting relaxes FOR such that we model the probability of an agent seeing another agent as a function of the distance between them (and this distribution defines what agents are ``viewable'').
We introduce a new $Q$-learning algorithm for both settings, addressing the \textbf{Partially Observable Mean Field (POMF) $Q$-learning} problem, using Bayesian updates to maintain a distribution over the mean action parameter. 

This paper's contributions are to (1) introduce two novel POMF settings, (2) introduce two novel algorithms for these settings, (3) prove that the first algorithm ends up close to the Nash $Q$-value \cite{hu2003nash}, and (4) empirically show that both algorithms outperform existing baselines in three complex, large-scale tasks. We will assume stationary strategies as do other related previous work \cite{hu2003nash} \cite{pmlr-v80-yang18d}. 
Our full paper with appendices is available on arXiv \cite{subramanian2020partially}.

\section{Background Concepts}
\textbf{Reinforcement learning} \cite{sutton2018reinforcement} is a problem formulated on top of MDPs $\langle S,A,P,R \rangle$, where $S$ is the state space that contains the environmental information accessible to an agent at each time step, $A$ gives the actions that the agent can take at each time step, the reward function $R$ provides real-valued rewards at each time step, and the transition dynamics $P$ is the probability of moving to a state $s'$ when the agent takes action $a$ at state $s$. 
$Q$-learning \cite{watkins1992q} learns a policy ($S \mapsto A$) by updating $Q$-values based on experience.

\textbf{Stochastic games} generalize from single agent to $N$-agent MDPs. Each step in a stochastic game (a stage game) depends on the experiences of the agents in previous stages. A $N$-player stochastic game is defined as a tuple $\langle S, A^1, \ldots, A^N, R^1, \ldots, R^N , P, \gamma \rangle$, 
where $S$ is the state space, $A^j$ is the action space of agent $j$, and $R^j: S \times A^1 \cdots \times A^N \rightarrow \Re$ is the reward function of agent $j$. Agents maximize their discounted sum of rewards with $\gamma \in [0,1)$ as the discount factor. From this formulation, it can be seen that agents can have completely different reward functions (competitive and competitive-cooperative games) or can agree to maintain a shared reward structure (cooperative games). Transition function $p: S \times A^1 \cdots \times A^N \rightarrow \Omega(S)$, returns the probability distribution over the next state ($\Omega(S)$) when the system transitions from state $s$ given actions $(a^1, \ldots, a^N)$ for all agents. The joint action is $\boldsymbol{a} \delequal [a^1, \ldots, a^N]$. The transition probabilities are assumed to satisfy $\sum_{s'} p(s'|s,a^1, \cdots, a^N) = 1$. The joint policy (strategy) of agents can be denoted by $\boldsymbol{\pi} \delequal [\pi^1, \ldots, \pi^N]$. Given an initial state $s$, the value function of agent $j$ is the expected cumulative discounted reward given by  $
    v^j_{\boldsymbol{\pi}}(s) = \sum_{t=0}^\infty \gamma^t \E_{\boldsymbol{\pi}, p} [r^j_t|s_0 = s]
$. The $Q$-function can then be formulated as $Q^j_\pi(s,\boldsymbol{a}) = r^j(s, \boldsymbol{a}) + \gamma \E_{s' \sim p}[v^j_\pi(s')]$, where $s'$ represents the next state.

\textbf{Nash Q-learning:} Hu and Wellman \cite{hu2003nash} extended the Nash equilibrium solution concept in game theory to stochastic games. The Nash equilibrium of a general sum stochastic game is defined as a tuple of strategies $(\pi^1_*, \cdots, \pi^N_*)$, such that for all $s \in S$ \cite{hu2003nash},
$$v^i(s, \pi^1_*, \cdots,\pi^i_*, \cdots, \pi^N_*) \geq v^i(s, \pi^1_*, \cdots, \pi^i, \cdots, \pi^N_* )
         \quad
         \forall \pi^i \in \Pi^i
$$
\noindent Here, $v^i$ denotes the value function of agent $i$. This implies that no agent can deviate from its equilibrium strategy and get a strictly higher payoff when all other agents are playing their equilibrium strategies. 
The Nash $Q$-function, $Q^i_*(s, \boldsymbol{a})$, is the sum of agent $i$'s current reward and its discounted future rewards when all agents follow the Nash equilibrium strategy. Hu and Wellman proved that under a set of assumptions, the Nash operator defined by 
$
\mathscr{H}^{\textrm{Nash}}\boldsymbol{Q}(s,\boldsymbol{a}) = \E_{s' \sim p} [\boldsymbol{r}(s,\boldsymbol{a}) + \gamma \boldsymbol{v}^{\textrm{Nash}}(s')]
$
converges to the $Q$ value of the Nash equilibrium. Here, $\boldsymbol{Q} \delequal [Q^1, \ldots, Q^N]$,  $\boldsymbol{r}(s, \boldsymbol{a}) = [r^1(s,\boldsymbol{a}), \ldots,  r^N(s, \boldsymbol{a})]$ and $\boldsymbol{v}^{Nash}(s) \delequal [v^1_{\boldsymbol{\pi}^*}(s), \ldots, v^N_{\boldsymbol{\pi}^*}(s)]$.

\textbf{Mean field reinforcement learning} extends the stochastic game framework to environments where the number of agents are infinite in the limit \cite{lasry2007mean}. All agents are assumed to be indistinguishable and independent from each other. In this case, all the agents in the environment can be approximated as a single virtual agent to which the learning agent (called the \emph{central agent}) formulates best response strategies. Yang et al.\ \cite{pmlr-v80-yang18d} approximates the multi-agent $Q$-function by the mean field $Q$-function (MFQ) using an additive decomposition and Taylor's expansion (Eq.~\ref{eq:approxQ}). 
\begin{equation}
Q^{j}(s_t,\textbf{a}_t) \approx Q^j(s_t, a^j_t, \overline{a}_t^j)\label{eq:approxQ}
\end{equation}
The MFQ is recurrently updated using Eqs.~\ref{eq:MFQ} -- \ref{eq:updatepolicy}:
\begin{gather} 
Q^j(s_t,a^j_t, \overline{a}^j_{t})  =  (1-\alpha) Q^j(s_t,a^j_t, \overline{a}^j_{t}) + \alpha[r^j_t + \gamma v^j(s_{t+1})] \label{eq:MFQ}
\end{gather}
\begin{align}
& \mbox{where } v^{j}(s_{t+1}) = \sum_{a^j_{t+1}}\pi^j(a^j_{t+1}|s_{t+1},\overline{a}^j_{t})  Q^j(s_{t+1},a^j_{t+1}, \overline{a}^j_{t}) \label{eq:yangvalueupdate} \\
& %\mbox{where } 
\overline{a}^j_{t} = \frac{1}{N^j} \sum_{k \neq j} a^k_t, a^k_t \sim \pi^k(\cdot|s_t,\overline{a}^k_{t-1}) \label{eq:updatemeana} \\
& \mbox{and } \pi^j(a^j_t|s_t, \overline{a}^j_{t-1}) = \frac{\exp(-\beta Q^j(s_t,a^j_t, \overline{a}^j_{t-1}))}{\sum_{a^{j'}_t\in A^j}\exp(-\beta Q^j(s_t,a^{j'}_t,\overline{a}^{j}_{t-1}))} \label{eq:updatepolicy}
\end{align}
$r^j_t$ is the reward for agent $j$, $s_t$ is the (global) old state, $s_{t+1}$ is the (global) resulting state, $\alpha$ is the learning rate, $v^j$ is the value function of $j$ and $\beta$ is the Boltzmann parameter. $a^j_t$ denotes the (discrete) action of agent $j$ represented as a one-hot encoding whose components are one of the actions in the action space. The $\overline{a}^j_t$ is the mean action of all other agents apart from $j$ and $\pi$ denotes the Boltzmann policy. In Eq.~\ref{eq:yangvalueupdate}, there is no expectation over $\overline{a}^j$, because Yang et al.~\cite{pmlr-v80-yang18d} guaranteed that the MFQ updates will be greedy in the limit ($t \xrightarrow{} \infty$). Finally, $N^j$ is the number of agents in the neighbourhood of $j$. We highlight that, for the mean action calculation in Eq. \ref{eq:updatemeana}, the policies of all other agents needs to be maintained by the central agent. Now, this policy can only be obtained by observing all other agents at every time step, which is a strong assumption in a large environment with many agents. Yang et al., overcome this problem by introducing \emph{neighbourhoods}. However, the neighbourhood needs to be large enough to contain the whole environment for these methods to work, as agents can go in and out of the neighbourhoods and go out of vicinity otherwise, which will make the computation of mean action as in Eq. \ref{eq:updatemeana}  inapplicable. In our work, we will relax this strong assumption in estimating the mean field action. We will not assume the observability of all other agents.

\section{Partially Observable Mean Field Q-Learning: FOR}\label{sec:POMFQFOR}

In this section, we study the Fixed Observation Radius (FOR) version of our problem, where all agents within a fixed neighbourhood from the central agent are visible to the central agent, and the others are not visible.
Our setting is same as that in Yang et al.\ \cite{pmlr-v80-yang18d} but we proceed to relax the assumption of global state observability.
We modify the update in Eqs. \ref{eq:MFQ} -- \ref{eq:updatepolicy} by maintaining a categorical distribution for the mean action parameter ($\overline{a}$). We will only use the local state $s^j$ of agent $j$ and not the global state. Eq. \ref{eq:qupdate} gives our corresponding $Q$ update equation. Since the conjugate prior of a categorical distribution is the Dirichlet distribution, we use a Dirichlet prior for this parameter. Let $L$ be the size of the action space. 
Let $\eta$ denote the parameters of the Dirichlet $(\eta_1, \ldots, \eta_L)$, $\mathcal{\theta}$ denote a categorical distribution $(\theta_1, \ldots, \theta_L)$, and $\mathcal{X}$ denote an observed action sample $(x_1, \ldots, x_G)$ of $G$ agents. Then the Dirichlet for agent $j$ can be given by $
    \mathcal{D}^j(\mathcal{\theta} | \eta) \propto \theta_1^{\eta_1 - 1} \cdots \theta_L^{\eta_L - 1}$ and the likelihood is given by $
    p(\mathcal{X} | \theta) \propto \theta_1^{[\mathcal{X} = 1]} \cdots \theta_L^{[\mathcal{X} = L]}
    \propto
     \theta_1^{c_1} \cdots \theta_L^{c_L}
$, where $[X = i]$ is the Iverson bracket, which evaluates to 1 if $X = i$ and $0$ otherwise. This value corresponds to the number of occurrences of each category ($c_1, \ldots, c_L$), denoted by $c$. Using a Bayesian update, the posterior is a Dirichlet distribution given by Eq.\  \ref{eq:bayesianupdate} where the parameters of this Dirichlet are given by $\mathcal{D}^j(\theta|\eta + c)$. 
 
 The modified $Q$ updates are: 
\begin{gather}
 Q^j(s^j_t,a^j_t, \Tilde{a}^j_t)  =  (1-\alpha) Q^j(s^j_t,a^j_t, \Tilde{a}^j_t) + \alpha[r^j_t + \gamma v(s^j_{t+1})] \label{eq:qupdate} \\
 \mathcal{D}^j(\mathcal{\theta}) \propto \theta_1^{\eta_1 - 1 + c_1} \cdots \theta_L^{\eta_L - 1 + c_L} ; \quad \mathcal{D}^j(\theta|\mathcal{\eta} + c) \label{eq:bayesianupdate}
 \end{gather}
 \begin{align}
 & \mbox{Where }    v^j(s^j_{t+1}) = \sum_{a^j_{t+1}}\pi^j(a^j_{t+1}|s^j_{t+1},\Tilde{a}^j_t)   Q^j(s^j_{t+1},a^j_{t+1}, \Tilde{a}^j_t) \label{eq:valueupdate}\\
& \mbox{ }          \Tilde{a}^j_{i,t} \sim \mathcal{D}^j(\theta; \eta + c); \quad \Tilde{a}^j _t = \frac{1}{\mathcal{S}}\sum_{i=1}^{i=\mathcal{S}} \Tilde{a}^j_{i,t} \label{eq:meanaupdate}\\
 & \mbox{and }    \pi^j(a^j_t|s^j_t, \Tilde{a}^j_{t-1}) = \frac{\exp(-\beta Q^j(s^j_t,a^j_t, \Tilde{a}^j_{t-1}))}{\sum_{a^{j'}_t\in A^j}\exp(-\beta Q^j(s^j_t,a^{j'}_t,\Tilde{a}^{j}_{t-1}))}
     \label{eq:policyupdate}
\end{align}
We have replaced the mean field aggregation from Yang et al.\ (Eq.\ \ref{eq:updatemeana}) with the Bayesian updates of the Dirichlet distribution from Eq.\  \ref{eq:bayesianupdate} and we take $\mathcal{S}$ samples from this distribution in Eq.\  \ref{eq:meanaupdate} to estimate the partially observable mean action ($\Tilde{a}$). This approach relaxes the assumption of complete observability of the global state. We use samples from the Dirichlet to introduce noise in the mean action parameter, enabling further exploration and helping agents to escape local optima. Being stuck in a local optimum is one of the major reasons for poor performance of learning algorithms in large scale systems. For example, Guo et al.~\cite{guo2019learning} show that MFQ and Independent $Q$-learning (IL) \cite{tan1993multi} remain stuck at a local optimum and do not move towards a global optimum in many settings, even after many training episodes. Yang et al.\ also report that the Mean Field Actor Critic (MFAC) and MFQ algorithms may remain stuck at a local optimum for a long period of training episodes in a simple Gaussian squeeze environment as the number of agents becomes exponentially large. Intuitively, this problem is even worse in a partially observable setting as the agents get a smaller observation and their best response policy is directed towards this observation sample. 
Sampling methods as in Eq. \ref{eq:meanaupdate} are also used in established algorithms like Thompson sampling \cite{thompson1933likelihood,osband2016deep}.
Finally, we update the Boltzmann policy like Yang et al.\ in Eq. \ref{eq:policyupdate}. We provide more theoretical guarantees for our update equations in Section \ref{sec:theoreticalresults}. 

This version of our problem is generally applicable to many different environments. However, in some domains, agents may not be able to see all the other agents in the vicinity, but closer agents will have a high probability of being seen. 
The next section considers a new version of our problem where some special kinds of distributions are used to model the observed agents. 
% We introduce an algorithm where explicit knowledge of these distributions are exploited to provide a more suitable approach in such environments. 

\section{Partially Observable Mean Field Q-Learning: PDO}\label{sec:PDO}

This section considers the Probabilistic Distance-based Observability (PDO) problem, assuming that each agent can observe other agents with some probability that decreases as distance increases.  

We introduce a distance vector $\mathscr{D}$ that represents the distance of other agents in the environment to the central agent. Hence, $\mathscr{D} = (d_1, \ldots, d_N)$, where $d_i$ denotes the distance of agent $i$ from the central agent. We use the exponential distribution to model the probability of the distance of agent $i$ from the central agent. Exponential distribution assigns a higher probability to smaller distances and this probability exponentially drops off as distance increases. Since, in a large environment the agents that matter are closer to the central agent, than far off, the exponential distribution is appropriate to model this variable. We drop the subscript of $d$ for clarity. This distribution is parameterized by $\hat{\theta}$ so that $d| \hat{\theta} \propto \exp(\hat{\theta})$. Since the conjugate prior of the exponential distribution is the gamma distribution, we use a gamma prior, and the prior distribution is parameterised by $\alpha, \beta$. Hence, we write $
\hat{\theta} \propto Gamma(\hat{\alpha}, \hat{\beta})
$.

We also maintain an additional parameter $b_i$ that determines whether a given agent $i$ is visible to the central agent $j$. The variable $b_i$ takes two values: 1 if this agent is in the field of view and 0 if this agent is not in the field of view. Again, we will drop the subscript of $b$. We maintain a Bernoulli distribution conditioned on the distance $d$.  The probability that an agent at a distance $d$ is visible is given by $Pr(b=1|d,\lambda)=\lambda e^{-d\lambda}$.  Note that this is not an exponential distribution, but rather a Bernoulli distribution with a probability defined by the same algebraic formula as the exponential distribution. In this setting, we will assume that the central agent will see varying numbers of other agents based on this distribution. Since the parameter $\lambda$ cannot be estimated by an agent from observation (the agent needs to know which other agents it is seeing and not seeing to infer $\lambda$), we will assume that the scalar value of $\lambda$ is common knowledge for all the agents. Since $\lambda e^{-\lambda d}$ should be in $[0,1]$ because it is a probability, only $\lambda$ values in $[0,1]$ satisfy this requirement. We will fix the value of $\lambda$ to be 1, but it could be any other value in the given range. This gives a definite distribution that determines the conditional of $b$.
We are particularly interested in the posterior term $Pr(\hat{\theta} | d,b=1)$, which denotes the probability of $\hat{\theta}$, given the distance $d$ of another agent $i$, and that $i$ is in the field of view of the central agent $j$. 
    \begin{flalign}
         Pr(\hat{\theta}|d,b=1) \propto Pr(d|\hat{\theta},b=1)Pr(\hat{\theta}|b=1) \propto Pr(d|\hat{\theta},b=1)Pr(\hat{\theta})
         \label{eq:probability1}
    \end{flalign}
In the last term of Eq.\  \ref{eq:probability1}, the variable $\theta$ does not depend on the variable $b$, so we remove the conditional. Now consider, 

\begin{flalign}
    & Pr(d|\hat{\theta}, b=1) = Pr(d,b|\hat{\theta}) / Pr(b|\hat{\theta})
   \nonumber \\
   & = Pr(b=1|\hat{\theta},d)Pr(d|\hat{\theta}) / Pr(b=1|\hat{\theta})
    \nonumber \\
   & = Pr(b=1|d)Pr(d|\hat{\theta}) / Pr(b=1|\hat{\theta})
  \nonumber \\
   & = \lambda e^{-d} \lambda \hat{\theta} e^{-d \hat{\theta}} / \int_d Pr(b=1|d) Pr(d|\hat{\theta})
    \nonumber \\
  &  = e^{-d} \hat{\theta} e^{-d \hat{\theta}} / \int^{d=\infty}_{d=0} \hat{\theta} e^{-d \hat{\theta}} \lambda e^{-d \lambda} 
  = e^{-d} \hat{\theta} e^{-d \hat{\theta}} / \int^{d=\infty}_{d=0} \hat{\theta} e^{-d (\hat{\theta} + 1)}
   \nonumber \\
   & = e^{-d} \hat{\theta} e^{-d \hat{\theta}} (\hat{\theta} + 1) / \hat{\theta}
    \label{eq:probability2} 
\end{flalign}  
Applying Eq.\  \ref{eq:probability2} in Eq.\  \ref{eq:probability1}, 
\begin{equation*}
    \begin{array}{l}
   Pr(\hat{\theta}|d,b=1) \propto Gamma(\hat{\alpha}, \hat{\beta}) \times  e^{-d} \hat{\theta} e^{-d \hat{\theta}} (\hat{\theta} + 1) / \hat{\theta}
   \\
    \propto \hat{\theta}^{\hat{\alpha}-1} e^{-\hat{\beta} \hat{\theta}} \times
    e^{-d}\hat{\theta} e^{-d \hat{\theta}} (\hat{\theta} + 1) / \hat{\theta}
    \\
  \propto \hat{\theta}^{\hat{\alpha}} e^{-\hat{\theta}(\hat{\beta} + d - d/\hat{\theta})}  + \hat{\theta}^{\hat{\alpha}-1} e^{-\hat{\theta}(\hat{\beta} + d - d/\hat{\theta})} 
    \end{array}
\end{equation*}
The posterior of $\hat{\theta}$ is therefore given by a mixture of Gamma distributions (i.e., $Gamma(\hat{\alpha} + 1, d + \hat{\beta} - d/\hat{\theta})$ and $Gamma(\hat{\alpha}, d + \hat{\beta} - d/\hat{\theta})$).
We can obtain a single Gamma posterior, corresponding to a projection of this mixture of Gammas, by updating the new value of $\hat{\alpha} $ as $\hat{\alpha} + 0.5$. We sample from this single Gamma distribution to get a new parameter $\overline{\lambda}$ (Eq.\  \ref{eq:distanceupdate}). We denote the Gamma distribution for the agent $j$ using superscript $j$ (Eq. \ \ref{eq:posteriorupdate}). Hence, we get:

\begin{gather}
Q^j(s^j_t,a^j_t, \Tilde{a}^j_t, \overline{\lambda}^j_t)  =  (1-\alpha) Q^j(s^j_t,a^j_t, \Tilde{a}^j_t, \overline{\lambda}^j_t) + \alpha[r^j_t + \gamma v^j(s^j_{t+1})] \label{eq:pdoqfunction}\\
\hat{\theta}_{P} \propto Gamma^j(\hat{\alpha} + 1, d + \hat{\beta} - d/\hat{\theta}) + Gamma^j(\hat{\alpha}, d + \hat{\beta} - d/\hat{\theta})
\label{eq:posteriorupdate}
\end{gather}
Where:
\begin{align}
 \overline{\lambda}^j_{i,t} \sim [Gamma^j(\hat{\alpha} + 0.5, d + \hat{\beta}  - d/\hat{\theta})];\quad\overline{\lambda}_{t}^j = \frac{1}{\mathcal{G}}\sum_{i=1}^{i=\mathcal{G}}  \overline{\lambda}^j_{i,t} \label{eq:distanceupdate}\\
v^j(s^j_{t+1}) = \sum_{a^j_{t+1}}\pi^j(a^j_{t+1}|s^j_{t+1},\Tilde{a}^j_t, \overline{\lambda}_t) Q^j(s^j_{t+1},a^j_{t+1}, \Tilde{a}^j_t, \overline{\lambda}^j_t) \label{eq:valueupdate2}
\end{align}
\begin{align}
\pi^j(a^j_t|s^j_t, \Tilde{a}^j_{t-1}, \overline{\lambda}^j_{t-1}) &= \frac{\exp(-\beta Q^j(s^j_t,a^j_t, \Tilde{a}^j_{t-1}, \overline{\lambda}^j_{t-1}))}{\displaystyle\sum_{a^{j'}_t\in A^j}\exp(-\beta Q^j(s^j_t,a^{j'}_t,\Tilde{a}^{j}_{t-1}, \overline{\lambda}^j_{t-1}))} \label{eq:policy2update}
\end{align}

All the variables above have the same meaning as in Eqs.\ \ref{eq:qupdate} -- \ref{eq:policyupdate}. The estimate of the $\Tilde{a}$ is obtained as in Eq.\  \ref{eq:meanaupdate}. The $\hat{\theta}_{P}$ denotes the new (posterior) value of $\hat{\theta}$. The $\overline{\lambda}$ parameter is updated by sampling from the Gamma distribution, as in Eq.\  \ref{eq:distanceupdate}, by taking $\mathcal{G}$ samples.

\section{Algorithm Implementations}

The implementation of POMFQ follows prior work \cite{pmlr-v80-yang18d} that uses neural networks --- $Q$-functions are parameterized using weights $\phi$, but tabular representations or other function approximators should also work. 
Our algorithms are an integration of the respective update equations with Deep $Q$-learning (DQN) \cite{mnih2015human}. 
Algorithm \ref{alg:partialQV1} gives pseudo code for the algorithm for the ``FOR'' case and Algorithm \ref{alg:partialQV2} for the ``PDO'' case. The lines in Algorithm \ref{alg:partialQV2} that have changed from Algorithm \ref{alg:partialQV1} are marked in blue. We provide a complexity analysis of our algorithms in Appendix \ref{sec:complexityanalysisappendix}.

\begin{algorithm}[ht]
\fontsize{7pt}{7pt}\selectfont
\caption{Partially observable mean field $Q$ Learning - FOR}
\label{alg:partialQV1}
\begin{algorithmic}[1] %[1] enables line numbers

\STATE Initialize the weights of $Q$-functions $Q_{\phi^j}, Q_{\phi^j_{\_}}$ for all agents $j \in {1, \ldots, N}$. 
\STATE Initialize the Dirichlet parameter $\mathcal{D}^j(\theta)$ for all agents $j$.
\STATE Initialize the mean action $\overline{a}^j$ for each agent $j \in {1, \ldots, N}$.  
\STATE Initialize the total steps (T) and total episodes (E). 
\WHILE {Episode $<$ E}
\WHILE{Step $<$ T}
\STATE For each agent $j$, sample $a^j$ from the policy induced by $Q_{\phi^j}$ according to Eq.~\ref{eq:policyupdate} with the current mean action $\Tilde{a}^{j}$ and the exploration rate $\beta$. 
\STATE For each agent $j$, update its  Dirichlet distribution (Eq.~\ref{eq:bayesianupdate}).
\STATE For each agent $j$, compute the new mean action  $\Tilde{a}^{j}$ (Eq. \ref{eq:meanaupdate}).
\STATE Execute the joint action $\textbf{a} = [a^1,\ldots, a^N]$. Observe the rewards $\textbf{r} = [r^1,\ldots, r^N]$ and the next state $\textbf{s'} = [s'^1, \ldots, s'^N]$. 
\STATE Store $\langle \textbf{s},\textbf{a},\textbf{r},\textbf{s'},\Tilde{\textbf{a}}\rangle$ in replay buffer $B$, where $\Tilde{\textbf{a}}{=}[\Tilde{a}^1, \ldots, \Tilde{a}^N]$ is the mean action. 
\ENDWHILE
\WHILE{$j$ = $1$ to $N$}
\STATE Sample a minibatch of K experiences  $  \langle \textbf{s},\textbf{a},\textbf{r},\textbf{s'},\Tilde{\textbf{a}}   \rangle$
from $B$.
\STATE Set $y^j = r^j + \gamma v^{POMF}_{\phi^j_{\_}}(s')$ according to Eq. \ref{eq:valueupdate}.
\STATE Update Q network by minimizing the loss $L(\phi^j) {=} \frac{1}{k} \sum (y^j - Q_{\phi^j}(s^j, a^j, \Tilde{a}^{j}))^2$.
\ENDWHILE
\STATE Update params of target network for each agent $j$: $\phi^{j}_{\_} \leftarrow \tau \phi^j + (1 - \tau) \phi^j_{\_}$.
\ENDWHILE
\end{algorithmic}
\end{algorithm}

\begin{algorithm}[h!]
\fontsize{7pt}{7pt}\selectfont
\caption{Partially observable mean field $Q$ Learning - PDO}
\label{alg:partialQV2}
\begin{algorithmic}[1] %[1] enables line numbers

\STATE Initialize the weights of $Q$ functions $Q_{\phi^j}, Q_{\phi^j_{\_}}$ for all agents $j \in {1, \ldots, N}$. 
\STATE Initialize the Dirichlet parameter $\theta$ in $\mathcal{D}^j(\theta)$ for all agents $j$.
\STATE \textcolor{blue}{Initialize $\hat{\alpha}$ and $\hat{\beta}$ in $Gamma^j(\hat{\alpha}, \hat{\beta})$ for all agents $j \in {1, \ldots, N}$.} 
\STATE Initialize the mean action $\overline{a}^j$ for each agent $j \in {1, \ldots, N}$.  
\STATE Initialize the total steps (T) and total episodes (E). 
\WHILE {Episode $<$ E}
\WHILE{Step $<$ T}
\STATE For each agent $j$ sample $a^j$ from the policy induced by $Q_{\phi^j}$ according to Eq. \ref{eq:policy2update} with the current mean action $\Tilde{a}^{j}$ and the exploration rate $\beta$.
\STATE For each agent $j$ update its Dirichlet distribution (Eq.\  \ref{eq:bayesianupdate}).
\STATE \textcolor{blue}{For each agent $j$, update its Gamma distribution (Eq.\  \ref{eq:posteriorupdate}).}
\STATE For each agent $j$, compute the new mean action  $\Tilde{a}^{j}$ (Eq. \ref{eq:meanaupdate}).
\STATE \textcolor{blue}{For each agent j, update parameter $\overline{\lambda}$ (Eq. \ref{eq:distanceupdate}).}
\STATE Execute the joint action $\textbf{a} = [a^1,\ldots ,a^N]$. Observe the rewards $\textbf{r} = [r^1, \ldots, r^N]$ and the next state $\textbf{s'} = [s'^1, \ldots, s'^N]$. 
\STATE \textcolor{blue}{Store $\langle \textbf{s},\textbf{a},\textbf{r},\textbf{s'},\Tilde{\textbf{a}} , \overline{\boldsymbol{\lambda}}\rangle$ in replay buffer $B$, s.t.~$\Tilde{\textbf{a}}{=}[\Tilde{a}^1, ..., \Tilde{a}^N]$, $\boldsymbol{\overline{\lambda}}{=} [\overline{\lambda}^1, ..., \overline{\lambda}^N]$}
\ENDWHILE
\WHILE{$j$ = $1$ to $N$}
\STATE \textcolor{blue}{Sample minibatch of K experiences  $  \langle \textbf{s},\textbf{a},\textbf{r},\textbf{s'},\Tilde{\textbf{a}}, \boldsymbol{{\overline{\lambda}}}   \rangle$
from $B$.}
\STATE \textcolor{blue}{Set $y^j = r^j + \gamma v^{POMF}_{\phi^j_{\_}}(s')$ according to Eq. \ref{eq:valueupdate2}}.
\STATE \textcolor{blue}{Update Q network by minimizing $L(\phi^j) {=} \frac{1}{k} \sum (y^j {-} Q_{\phi^j}(s^j, a^j, \overline{a}^{j}, \overline{\lambda}))^2$}.
\ENDWHILE
\STATE Update params of target network for each agent $j$: $\phi^{j}_{\_} \leftarrow \tau \phi^j + (1 - \tau) \phi^j_{\_}$.
\ENDWHILE
\end{algorithmic}
\end{algorithm}

\section{Theoretical Results}\label{sec:theoreticalresults}

The goal of this section is to show that our FOR $Q$-updates are guaranteed to converge to the Nash $Q$-value. We will begin by providing a technical result that is generally applicable for any stochastic processes of which the $Q$-function is a specific example. Then we have a sequence of theorems that lead us to bound the difference between the POMF $Q$-value and the Nash $Q$-value in the limit ($t \xrightarrow{} \infty$). We outline a number of common assumptions that are needed to prove these theorems. For the purposes of a direct comparison of the POMF $Q$-function and the Nash $Q$-function, we assume that we have a system of $N$ agents where agents have the full global state available and thus have the ability to perform a MFQ update (Eqs. \ref{eq:MFQ} -- \ref{eq:updatepolicy}) or a POMFQ update (Eqs. \ref{eq:qupdate} -- \ref{eq:policyupdate}). By the definition of a Nash equilibrium, every agent should have the knowledge of every other agent's strategy. To recall, in a Nash equilibrium, no agent will have an incentive to unilaterally deviate, given the knowledge of other agent strategies. Our objective is also to make a direct comparison between the POMFQ update and the MFQ update and hence we will use the FOR setting algorithms of POMFQ update in the theoretical analysis as it is most directly related to MFQ. 
In this section, we will show that a representative agent $j$'s $Q$-value will remain at least within a small distance of the Nash $Q$-value in the limit ($t \xrightarrow{} \infty$) as it performs a POMFQ update, which tells us that, in the worst case, the agents stay very close to the Nash equilibrium. We have provided a proof sketch for all our theorems in this section while the complete versions of our proofs can be found in Appendix \ref{appendix:proofappendix}. In a mean field setting, the homogeneity of agents allows us to drop the agent index $j$ \cite{lasry2007mean} for the value and $Q$-function, which we adopt for clarity. Also, ``w.p.1'' represents ``with probability one''.

Consider an update equation of the following form (using the Tsitsiklis \cite{tsitsiklis1994asynchronous} formulation): 
\begin{equation}\label{Eq:stochasticprocess}
    \begin{array}{l}
         x_i(t+1) = x_i(t) + \alpha_i(t)(F_i(x^i(t)) - x_i(t) + w_i(t))
    \end{array}
\end{equation}
Here, $x(t)$ is the value of vector $x$ at time $t$ and $x_i(t)$ denotes its $i$th component. Let, $F$ be a mapping from $\mathscr{R}^n$ into itself. Let $F_1, \ldots, F_n$ : $\mathscr{R}^n \rightarrow \mathscr{R}$ be the component mappings of $F$, that is $F(x) = (F_1(x), \ldots, F_n(x))$ for all $x \in \mathscr{R}^n$. Also, $w_i(t)$ is a noise term, and $x^i(t)$ can be defined as  
$
         x^i(t) = (x_1(\tau_1^i(t)), \cdots, x_n(\tau^i_n(t))) 
$, where each $\tau^i_j(t)$ satisfies $0 \leq \tau^i_j(t) \leq t$. 

Next, we state some assumptions. The first three are the same as those in Tsitsiklis \cite{tsitsiklis1994asynchronous}, but we modify the fourth assumption. The first assumption guarantees that old information is eventually discarded with probability one. The second assumption is a measurability condition and the third assumption is the learning rate condition, both of which are common in RL \cite{szepesvari1999unified} \cite{tsitsiklis1994asynchronous}. The Assumption \ref{assump:monotone}, is a condition on the $F$ mapping, which is a weaker version than the fourth assumption in Tsitsiklis \cite{tsitsiklis1994asynchronous}. 

\begin{assumption}\label{assumption:oldinformation}
For any $i$ and $j$, $\lim_{t \rightarrow \infty} \tau^i_j(t) = \infty$ w.p.1. 
\end{assumption}

\begin{assumption}\label{assumption:assorted}

a) $x(0)$ is $\mathcal{F}(0)$-measurable
\\
b) For every $i$, $j$, and $t$, $w_i(t)$ is $\mathcal{F}(t + 1)$-measurable
\\
c) For every $i$, $j$, and $t$, $\alpha_i(t)$ and $\tau_j^i(t)$ are $\mathcal{F}(t)$-measurable \\
d) For every $i$ and $t$, we have $\E[w_i(t) | \mathcal{F}(t)] = 0$  \\
e) For deterministic constants $A$ and $B$,
$$
         \E[w_i^2(t) | \mathcal{F}(t) ] \leq A + B max_j max_{\tau \leq t} |x_j(\tau)|^2
$$     
\end{assumption}

\begin{assumption}\label{assumption:learningrate}
The learning rates satisfy $0 \leq \alpha_i(t) < 1$.
 \end{assumption}

\begin{assumption}\label{assump:monotone}
a) The mapping $F$ is monotone; that is, if $x \leq y$, then $F(x) \leq F(y)$
\\
b) The mapping $F$ is continuous
\\
c) In the limit ($t \xrightarrow{} \infty$), the mapping $F$ is bounded in an interval $[x^* - D$, $x^* + D]$, where $x^*$ is some arbitrary point
\\
d) If $e \in \mathcal{R}^n$ is the vector with all components equal to 1, and $p$ is a positive scalar then, $
    F(x) - pe \leq F(x-pe) \leq F(x + pe) \leq F(x) + pe$

\end{assumption}

Now, we will state our first theorem. Theorem \ref{theorem:tsitskilis2} is a technical result that we obtain by extending Theorem 2 in Tsitsiklis \cite{tsitsiklis1994asynchronous}. We will use this result to derive the main result in Theorem \ref{theorem:boundtheorem}.

\begin{theorem}\label{theorem:tsitskilis2}

A stochastic process of the form given in Eq.\  \ref{Eq:stochasticprocess} remains bounded in the range $[x^* - 2D, x^* + 2D]$ in the limit, if Assumptions \ref{assumption:oldinformation} --  \ref{assump:monotone} hold, and if the process is guaranteed not to diverge to infinity. $D$ is the bound on the $F$ mapping in Assumption \ref{assump:monotone}(c).

\end{theorem}

\begin{proofsk}
Since the stochastic process in Eq.\  \ref{Eq:stochasticprocess} is guaranteed to stay bounded (Assumption \ref{assump:monotone}(c)), one can find other processes that lower bounds and upper bounds this process. Let us assume that we can show that the process in Eq.\  \ref{Eq:stochasticprocess} always stays bounded by these two processes after some finite time $t$ (that is for all $t' \geq t$). Now, if we can prove that the process $A$ is upper bounded by a finite value, this value will be the upper bound of the process in Eq.\  \ref{Eq:stochasticprocess} after $t$ as well. Similarly, the lower bound of $L$ will be its lower bound (after time $t$). 
% In Appendix \ref{appendix:proofappendix} we show that we can define particular processes $A$ and $L$ that satisfy these conditions and that $A$ will be upper bounded by $x^* + 2D$ and $L$ will be lower bounded by $x^* - 2D$ by involving Assumptions \ref{assumption:oldinformation} -- \ref{assump:monotone}. 
\end{proofsk}

Now we state three more assumptions, as used earlier
\cite{pmlr-v80-yang18d}.

\begin{assumption}\label{assumption:rewardbound}
Each action-value pair is visited infinitely often and the reward stays bounded.
\end{assumption}

\begin{assumption}\label{assumption:GLIE}
The agents' policy is Greedy in the Limit with Infinite Exploration (GLIE). 
\end{assumption}

\begin{assumption}\label{assumption:globaloptimum}
The Nash equilibrium can be considered a global optimum or a saddle point in every stage game of the stochastic game.

\end{assumption}

Assumption \ref{assumption:rewardbound} is very common in RL \cite{szepesvari1999unified}. Assumption \ref{assumption:GLIE} is needed to ensure that the agents are rational \cite{pmlr-v80-yang18d} and this is satisfied for POMFQ as the Boltzmann policy is known to be GLIE \cite{singh2000convergence}. Assumption \ref{assumption:globaloptimum} has been adopted by previous researchers \cite{hu2003nash,pmlr-v80-yang18d}. Hu and Wellman \cite{hu2003nash} consider this to be a strong assumption, but they note that this assumption is needed to prove convergence in theory, even though it is not needed to observe convergence in practice. 

Let $\Tilde{a}_i$ be a component of vector $\Tilde{a}$ and $\overline{a}_i$ be a component of vector $\overline{a}$. Now, we make a comparison between mean actions of the MFQ update (Eq.\  \ref{eq:updatemeana}) and the POMFQ update (Eq.\  \ref{eq:meanaupdate}). 

\begin{theorem}\label{theorem:abound}
The MFQ mean action and the POMFQ mean action both satisfy 
\begin{equation*}
    | \Tilde{a}_{i,t} - \overline{a}_{i,t}| \leq \sqrt{\frac{1}{2n} \log \frac{2}{\delta}}
\end{equation*}
as time $t \xrightarrow{} \infty$, with probability $>= \delta$, where $n$ is the number of samples observed. $\Tilde{a}$ is the mean action as obtained from the Dirichlet in Eq.\  \ref{eq:meanaupdate} and $\overline{a}$ is the mean action in Eq.\  \ref{eq:updatemeana}. 
\end{theorem}

\begin{proofsk}
This theorem is an application of the Hoeffding's bound which provides a probabilistic bound for the difference between the sample mean and the the true mean of a distribution. As $\Tilde{a}$ is an empirical mean of the samples $n$ observed at each time step, the Hoeffding's bound is applied to obtain the result. 
\end{proofsk}

\begin{theorem}\label{theorem:POMFMFbound}
When the $Q$-function is Lipschitz continuous (with constant M) with respect to mean actions, then the POMF $Q$-function will satisfy the following relationship: 
\begin{equation} \label{eq:qfunctionboundwithmeana}
    |Q^{POMF}(s_t, a_t, \Tilde{a}_{t-1}) - Q^{MF} (s_t, a_t, \overline{a}_{t-1})| \leq M \times L \times \log \frac{2}{\delta}\times \frac{1}{2n}
\end{equation}
\noindent
as $t \xrightarrow{} \infty$ with probability $\geq (\delta)^{L-1}$, where $L=|A|$ and $n$ is the number of samples.  
\end{theorem}

\begin{proofsk}
Once we have the bound on the mean actions of POMFQ update and MFQ update as in Theorem \ref{theorem:abound}, with the assumption of Lipschitz continuity, a corresponding bound can be derived for the respective $Q$-functions too. This is done by applying the bound of the mean actions in the Lipschitz condition. 
\end{proofsk}

From Theorem \ref{theorem:POMFMFbound}, we can see that in a similar setting, the POMFQ updates will not see a significant degradation in performance as compared to the MFQ updates. The probability of this holding is inversely proportional to the size of the action space available to each agent. In Theorem \ref{theorem:POMFMFbound}, the bound is between two $Q$-functions with the same state and action, but with different mean actions. Let $Z =  M \times L \times \log \frac{2}{\delta}\times \frac{1}{2n} $ and from Theorem \ref{theorem:POMFMFbound}, $|Q^{POMF}(s_t, a_t, \Tilde{a}_{t-1}) - Q^{MF}(s, a_t, \overline{a}_{t-1})| \leq  Z$. 
Now, we would like to directly compare the value estimates of POMFQ and MFQ updates. Consider two different actions $a^j$ and $b^j$ for agent $j$. Under the assumption that the mean field $Q$-function is $K$-Lipschitz continuous with respect to actions,

\begin{equation}\label{eq:boundonQwithsamea}
    |Q^{MF}(s^j_t, a^j_t, \overline{a}^j_{t-1}) - Q^{MF} (s^j_t, b^j_t, \overline{a}^j_{t-1})| \leq K |a^j_t - b^j_t | \leq K\sqrt{2} 
\end{equation}
In the last step, we applied the fact that all components of $a^j$ and $b^j$ are less than or equal to 1 (a one hot encoding). Assume that the optimal action for $Q^{POMF}$ is $a^*$ and for $Q^{MF}$ is $b^*$. Now consider, 

\begin{align}
&|v^{POMF}(s_{t+1}) - v^{MF}(s_{t+1}) |  
\nonumber \\
&= | \max_{a_{t+1}} Q^{POMF}(s_{t+1}, a_{t+1}, \Tilde{a}_t) -
\max_{b_{t+1}} Q^{MF}(s_{t+1}, b_{t+1}, \overline{a}_t) | 
 \nonumber\\
&= |Q^{POMF}(s_{t+1}, a^*_{t+1}, \Tilde{a}_t) - Q^{MF}(s_{t+1}, a^*_{t+1}, \overline{a}_t) \nonumber\\ 
&+ Q^{MF}(s_{t+1}, a^*_{t+1}, \overline{a}_t) 
    - Q^{MF}(s_{t+1}, b^*_{t+1}, \overline{a}_t)|
    \leq Z + K\sqrt{2}  \delequal D
    \label{eq:valuebound}
\end{align}
In the first step we apply the fact that the Boltzmann policy will become greedy in the limit ($t \xrightarrow{} \infty$). 
The last step is coming from Eqs. \ref{eq:qfunctionboundwithmeana} and \ref{eq:boundonQwithsamea}. We also reiterate that the Lipschitz continuity assumptions on the $Q$-function are consistent with prior work \cite{pmlr-v80-yang18d}.

\begin{theorem}\label{theorem:boundtheorem}
When we update the $Q$ functions using the partially observable update rule in Eq.\  \ref{eq:qupdate}, the process
satisfies the condition in the limit ($t \rightarrow \infty$):
\begin{equation*}
    |Q^*(s_t, \boldsymbol{a}_t) - Q^{POMF}(s_t, a_t, \Tilde{a}_t)| \leq 2D
\end{equation*}
\noindent
when Assumptions \ref{assumption:learningrate}, \ref{assumption:rewardbound}, and \ref{assumption:globaloptimum} hold. Here $Q^*$ is the Nash Q-value and $D$ is the bound for value functions in Eq.\  \ref{eq:valuebound}. This holds with probability at least $\delta^{L-1}$, where $L=|A|$.   
\end{theorem}

\begin{proofsk}
This result is an application of Theorem \ref{theorem:tsitskilis2}, where we show that all the assumptions of Theorem \ref{theorem:tsitskilis2} are satisfied by the conditions in this theorem. 
\end{proofsk}

It is important to note that we need only three minor assumptions (Assumptions \ref{assumption:learningrate}, \ref{assumption:rewardbound}, and \ref{assumption:globaloptimum}) to hold for Theorem 4, 
which is our main theoretical result. Theorem \ref{theorem:boundtheorem} shows that the POMFQ updates stay very close to the Nash equilibrium in the limit ($t \xrightarrow{} \infty$). The lower bound on the probability of this is high for a small action space and low for a large action space. In a multi-agent setting, the $Q$-updates are in the form of POMFQ updates, and do not have the (intuitive) effect of having any fixed point as commonly seen in RL. Theorem \ref{theorem:boundtheorem} 
proves our update rule is very close to the Nash equilibrium, a stationary point for the stochastic game. Hence, the policy in Eq.\  \ref{eq:updatepolicy} is approximately close to this stationary point, which guarantees that it becomes (asymptotically) stationary in the limit ($t \xrightarrow{} \infty$).

The distance between the POMF $Q$-function and the Nash $Q$-function is inversely proportional to the number of samples from the Dirichlet ($n$). If the agent chooses to take a large number of samples, the POMF $Q$-estimate is very close to the Nash $Q$-estimate, but this may lead to a degradation in performance due to 
% This is because having no 
% additional 
having no additional exploratory noise as discussed in Section \ref{sec:POMFQFOR}. 
% may reduce the performance.
In MARL, the Nash equilibrium is not a guarantee of optimal performance, but only a fixed point guarantee. 
The (self-interested) agents would still take finite samples, for better performance. To balance the theory and performance, the value of $n$ should not be too high nor too low. 

Appendix \ref{sec:isingappendix} provides an experimental illustration of Theorem \ref{theorem:boundtheorem} in the Ising model, a mathematical model used to describe the magnetic spins of atomic particles. This model was also used in \cite{pmlr-v80-yang18d}. 
We show that the distance (error) of the POMF $Q$-function (tabular implementation of the FOR updates) from the Nash $Q$-function stays bounded after a finite number of episodes as in Theorem \ref{theorem:boundtheorem}.

\section{Experiments and Results}\label{sec:expertiments}

This section empirically demonstrates that
using POMFQ updates will result in better performance in a partially observable environment than when using the MFQ updates. All the code for the experiments is open sourced \cite{sourcecode}.

We design three cooperative-competitive games for each of the two problems (FOR and PDO) within the MAgent framework \cite{zheng2018magent} to serve as testbeds. We will provide the important elements of these experimental domains here, while the comprehensive details (including exact reward functions and hyperparameter values) are deferred to Appendix \ref{sec:experimentaldetailsappendix}.   For all games, we have a two stage process: training and faceoff (test). We consider four algorithms for all the games: MFQ, MFAC, IL, and POMFQ. In each stage, there are two groups of agents: group A and group B. Since the agents do not know what kind of opponents they will see in the faceoff stage, they train themselves against another group that plays the same algorithm in the training stage. Thus, in the training stage, each algorithm will train two networks (groups A and B). 
In the faceoff stage, groups trained by different algorithms fight against each other. Our formulation is consistent with past research using the MAgent framework \cite{Srirammtmfrl2020,pmlr-v80-yang18d}. We plot the rewards obtained by group A in each episode for the training stage (group B also shows similar trends --- our games are not zero sum) and the number of games won by each algorithm in the faceoff stage. For statistical significance, we report p-values of an unpaired 2-sided t-test for particular episodes in the training stage and a Fischer's exact test for the average performances in the faceoff stage. We treat p-values of less than 0.05 as statistically significant differences. The tests are usually conducted between POMFQ and next best performing algorithm in the final episode of training for the training results.

The Multibattle game has two groups of agents fight against each other. There are 25 agents in each group for a total of 50. Agents learn to cooperate within the group and compete across the group to win. We analyse both the FOR and PDO cases. In FOR, information about nearby agents is available, but agents further than 6 units are hidden.
In PDO, the game engine maintains a Bernoulli distribution of visibility of each agent from every other agent as discussed in Section \ref{sec:PDO}. Based on this probability, each agent in PDO could see different numbers of
other agents at each time step. 
MFQ and MFAC use a frequentist strategy where agents observed at each time step are aggregated (Eq.\  \ref{eq:updatemeana}) to obtain a mean action. We run 3000 episodes of training in FOR and 2000 episodes in PDO. Each episode has a maximum of 500 steps. For the faceoff, group A trained using the first algorithm and group B trained using the second algorithm fight against each other for 1000 games. We report all results as an average of 20 independent runs for both training and faceoff (with standard deviation). In our experiments, an average of 6 -- 8 agents out of 50 agents are visible to the central agent at a given time step (averaged over the length of the game). Note that we use 50 agents per game, more agents are used in previous research \cite{Srirammtmfrl2020,pmlr-v80-yang18d}. In our case, the ratio of agents seen vs.\ the total number of agents matters more than the simple absolute number of agents in the competition.

\begin{figure}
	\subfloat[FOR -  Train]{{\includegraphics[width=0.22\textwidth, height=4cm]{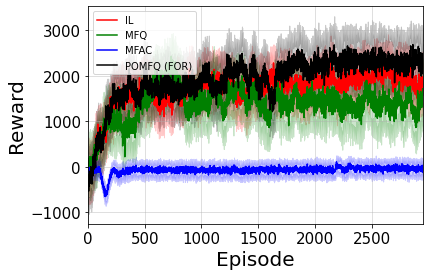} }}
	\subfloat[PDO -  Train]{{\includegraphics[width=0.22\textwidth, height=4cm]{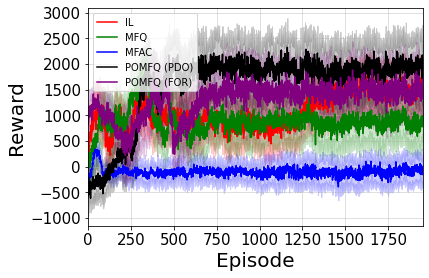} }}
	\\
	\subfloat[FOR -  Test]{{\includegraphics[width=0.22\textwidth, height=3.8cm]{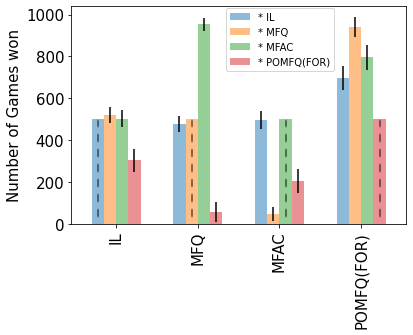} }}
	\subfloat[PDO -  Test]{{\includegraphics[width=0.22\textwidth, height=3.8cm]{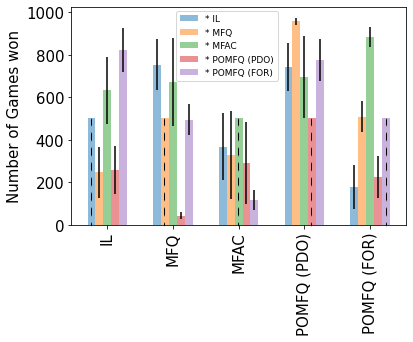} }}
  \caption{Multibattle results. The * in the legend of test plots denotes the opponent. For example, first orange bar (from the left) in the bar plots is result for IL.\ vs MFQ. The dashed lines indicate bars that we set for symmetry. We do not run faceoff experiments between the same algorithm. \vspace{-10pt} }%
	\label{fig:multibattle}
\end{figure}

In the FOR setting of the Multibattle domain (Figure \ref{fig:multibattle} (a)) the POMFQ algorithm plays the FOR variant (Algorithm \ref{alg:partialQV1}). POMFQ dominates other baselines from about 1800 episodes (p < 0.3) until the end (p < 0.03). We see that MFAC quickly falls into a poor local optimum and never recovers. The poor performance of MFAC in the MAgent games, compared to the other baselines, is consistent with previous work~\cite{pmlr-v80-yang18d, Srirammtmfrl2020}. 
Faceoff in the FOR case (Figure \ref{fig:multibattle}(c)) shows that POMFQ wins more than 50\% of the games against others (p < 0.01). 
An ablation study in Appendix \ref{sec:ablationstudyappendix} shows that performance improves with increase in viewing distance. 

In the PDO setting, we use both the FOR variant of POMFQ algorithm (no $\overline{\lambda}$ parameter) and the PDO variant of POMFQ algorithm (Algorithm \ref{alg:partialQV2}). We differentiate these two algorithms in the legends of Figures \ref{fig:multibattle}(b) and \ref{fig:multibattle}(d). The FOR variant loses out to the PDO algorithm that explicitly tracks the $\overline{\lambda}$ parameter (p < 0.02). If an algorithm bases decisions only on $\Tilde{a}$, as in Algorithm \ref{alg:partialQV1}, the agents do not know how risk seeking or risk averse their actions should be (when agents nearby are not visible). In this game, agents can choose to make an attack (risk seeking) or a move (risk averse). The additional parameter $\overline{\lambda}$ helps agents understand the uncertainty in not seeing some agents when making decisions. The PDO algorithm takes a lead over the other algorithms from roughly 900 episodes (p < 0.04) and maintains the lead until the end (p < 0.03). In faceoff, PDO wins more than 50\% (500) of the games against all other algorithms as seen in Figure \ref{fig:multibattle}(d) (p < 0.01).

\begin{figure}
	\subfloat[FOR - Train]{{\includegraphics[width=0.22\textwidth, height=4cm]{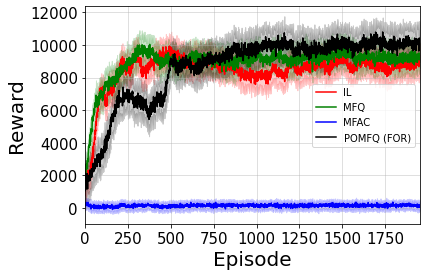} }}
	\subfloat[PDO - Train]{{\includegraphics[width=0.22\textwidth, height=4cm]{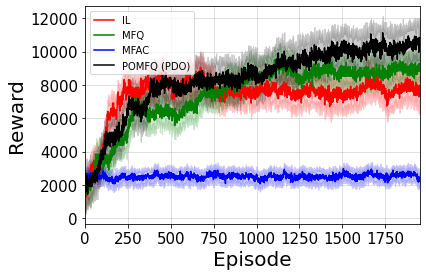} }}
	\\
	\subfloat[FOR -  Test]{{\includegraphics[width=0.22\textwidth, height=4cm]{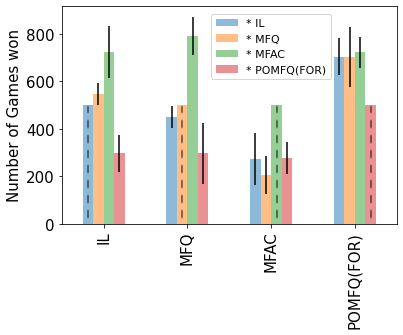} }}
	\subfloat[PDO - Test]{{\includegraphics[width=0.22\textwidth, height=4cm]{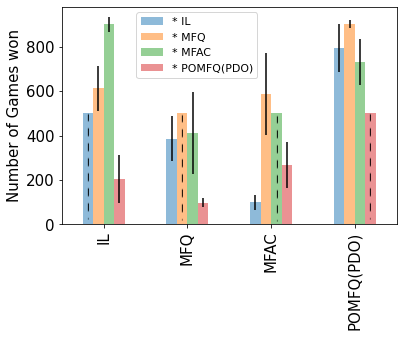} }}
  \caption{Training and faceoff results of Battle-Gathering game. 
  }%
	\label{fig:multigather}
\end{figure}

The second game, Battle-Gathering, is similar to the Multibattle game where a set of two groups of 50 agents are fighting against each other to win a battle, but with an addition of food resources scattered in the environment. All the agents are given an additional reward when capturing food, in addition to killing the competition (as in Multibattle). The training and faceoff are conducted similar to Multibattle game. Figure \ref{fig:multigather}(a) shows that the POMFQ algorithm 
dominates the other three algorithms from about 900 episodes (p < 0.03) till the end (p < 0.01). In the comparative battles (Figure \ref{fig:multigather}(c)), POMFQ has a clear lead over other algorithms (p < 0.01). MFQ and IL are similar in performance and MFAC loses to all other algorithms. We also observe this in the PDO domain train (p < 0.02, Figure \ref{fig:multigather}(b)) and test experiments (p < 0.01, Figure \ref{fig:multigather}(d)). 

\begin{figure}
	\subfloat[FOR - Train]{{\includegraphics[width=0.22\textwidth, height=4cm]{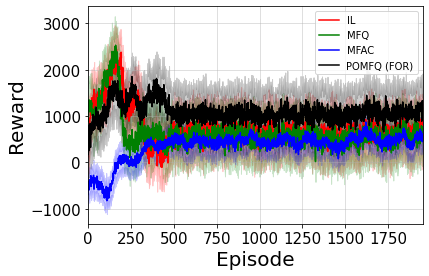} }}
	\subfloat[PDO - Train]{{\includegraphics[width=0.22\textwidth, height=4cm]{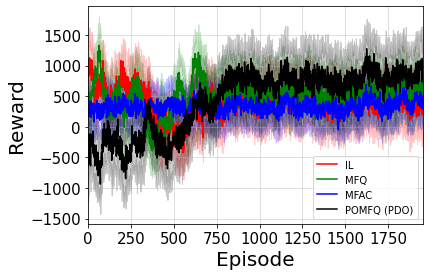} }}
	\\
	\subfloat[FOR - Test]{{\includegraphics[width=0.22\textwidth, height=4cm]{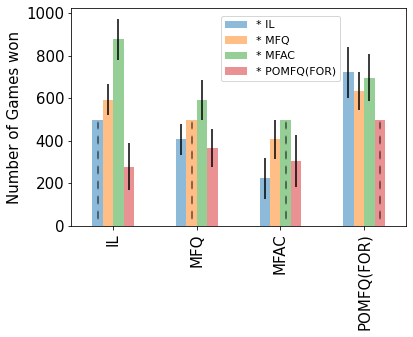} }}
	\subfloat[PDO -  Test]{{\includegraphics[width=0.22\textwidth, height=4cm]{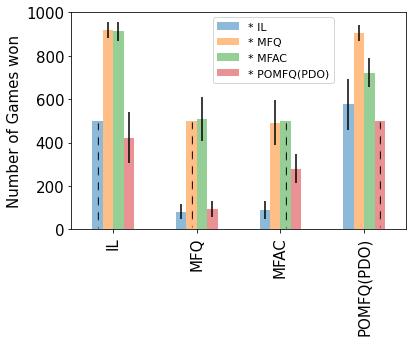} }}
  \caption{Training and faceoff results of Predator-Prey game. 
  }%
  
	\label{fig:predatorprey}
\end{figure}

The third game is a type of Predator-Prey domain, where there are two groups --- predators and prey. There are a total of 20 predator agents and 40 prey agents in our domain. The predators are stronger than the prey and have an objective of killing the prey. The prey are faster than the predator and try to escape from the predators. The training is conducted and rewards are plotted using the same procedure as in the Multibattle domain. Training performances are in Figures \ref{fig:predatorprey}(a) and \ref{fig:predatorprey}(c). The standard deviation of the performance in this game is considerably higher than the previous two games because we have two completely different groups that are trying to outperform each other in the environment. At different points in training, one team may have a higher performance than the other, and this lead can change over time.
In the first setting, Figure \ref{fig:predatorprey}(a), we can see that the POMFQ (FOR) shows a small lead over other baseline algorithms (at the end of training, p < 0.1). In the direct faceoff (Figure \ref{fig:predatorprey}(c)), POMFQ wins more games than the other algorithms (p < 0.01), 
 In the PDO setting too, the POMFQ-PDO algorithm has an edge over the others during the training phase (p < 0.4) and the testing (p < 0.01)(Figure \ref{fig:predatorprey}(b) and \ref{fig:predatorprey}(d)). 
As the p-values for the training suggest, POMFQ can be seen to have a better performance, but the results are not statistically significant. The faceoff results, on the other hand, are statistically significant (p < 0.01). We have run training for 2000 games and faceoff for 1000 games in the last two domains.

In three semantically different domains, we have shown that in the partially observable case, the MFQ and MFAC algorithms using frequentist strategies do not provide good performances. Also, the frequentist strategies (MFQ and MFAC) have worse performance in the harder PDO domain compared to the FOR domain. Sometimes, they also lose out to a simpler algorithm that does not even track the mean field parameter (IL). The FOR and PDO algorithms gives the best performance across both settings, as evidenced by the training and the test results. The training results clearly show that POMFQ never falls into a very poor local optimum like MFAC often does. The test results show that in a direct face-off, POMFQ outperforms all other algorithms.  The p-values indicate that our results are statistically significant. Additionally, in Appendix \ref{sec:recurrentbaselinesappendix}, we provide comparisons of the POMFQ - FOR and PDO algorithms with two more baselines, recurrent versions of IL and MFQ, in the same three MAgent domains and the results show that POMFQ has clear advantages compared to these recurrent baselines as well.

\section{Related Work}
Mean field games were introduced by Lasry and Lions \cite{lasry2007mean}, extending mean field theory \cite{ring1996relativistic, stanley1973introduction} to the stochastic games framework.  
The stochastic games formulation was obtained by extending MDPs to MARL environments  \cite{hu2003nash,littman1994markov}. 
Recent research has actively used the mean field games construct in a MARL setting, allowing tractable solutions in environments in which many agents participate. Model-based solutions have also been tried in this setting \cite{kizilkale2012mean}, but the model is specific to the application domain and these methods do not generalize well. Subramanian and Mahajan \cite{subramanian2019reinforcement} analyze the problem using a stationary mean field. In contrast to our approach, this paper needs strict assumptions regarding this stationarity, which do not hold in practice. Mguni et al.\ \cite{mguni2018decentralised} approaches this problem using the fictitious play technique. They provide strong theoretical properties for their algorithms, but only in the finite time horizon case. These results do not directly hold for infinite horizons. Additionally, strict assumptions on the reward function and fictitious property \cite{berger2007brown} assumption makes their algorithms less generally applicable. In fictitious play, each agent assumes that its opponents are playing stationary strategies. Thus, the response of each agent is a best response to the empirical frequency of their opponents. Another work by the same authors \cite{mguni2019coordinating} introduces an algorithm and provides theoretical analysis for the mean field learning problem in cooperative environments. Our methods, on the other hand, work for both cooperative and competitive domains. Along the same lines, the work by Elie et al.\ \cite{elie2019approximate,elie2020convergence} contributes fictitious play based techniques to solve mean field games with general theoretical properties based on quantifying the errors accumulated at each time step.  However, the strict assumptions on the reward function in addition to the fictitious play assumption is also present in the work by Elie et al. In our work, the agents do not make the fictitious play assumption for best responses. Yang et al.\ \cite{pmlr-v80-yang18d} do not have the limitations of other works noted here, but it assumes the global state is observable for all agents and a local action is taken from it. This has been relaxed by us.

\section{Conclusion}

This paper considers many agent RL problems where the exact cumulative metrics regarding the mean field behaviour is not available and only local information is available. We used two variants of this problem and provided practical algorithms that work in both settings. We empirically showed that our approach is better than previous methods that used a simple aggregate of neighbourhood agents to estimate the mean field action. We theoretically showed that POMFQ stays close to the Nash $Q$ under common assumptions.

In future work, we would like to relax some assumptions about the Bayesian approach using conjugate priors and make our analysis more generally applicable. Additionally, different observation distributions could allow the direction of view to determine the ``viewable'' agents, such as when agents in front of another agent are more likely to be seen than agents behind it.

\section{Acknowledgements}
Resources used in preparing this research were provided by the province of Ontario and the government of Canada through CIFAR, NSERC and companies sponsoring the Vector Institute. Part of this work has taken place in the Intelligent Robot Learning Lab at the University of Alberta, which is supported in part by research grants from the Alberta Machine Intelligence Institute (Amii), CIFAR, and NSERC.

\newpage
\clearpage

\bibliographystyle{ACM-Reference-Format} 
\bibliography{main}

\newpage

\appendix

\section{Proof for Theorems}\label{appendix:proofappendix}

All the theorems from the main paper are repeated here. The detailed proofs of these theorems are given. No new theorems are provided in this section.  

\begin{theorem2}

A stochastic process of the form given in Eq.\  \ref{Eq:stochasticprocess} remains bounded in the range $[x^* - 2D, x^* + 2D]$ in the limit, if Assumptions \ref{assumption:oldinformation} --  \ref{assump:monotone} hold, and if the process is guaranteed not to diverge to infinity. $D$ is the bound on the $F$ mapping in Assumption \ref{assump:monotone}(c).
\end{theorem2}
\begin{proof}

Let $p$ be an arbitrary scalar such that 
\begin{equation*}
    x^* - pe \leq x(t) \leq x^* + pe; \quad \forall t > t'. 
\end{equation*}

Such a $p$ is possible because the theorem assumes that the process $x(t)$ will not diverge to infinity. Here $t'$ is a point after which the mapping $F$ is bounded in $[x^* - D, x^* + D]$. We can find such a $t'$ with a high probability due to Assumption \ref{assump:monotone}(c). Our proof lies in the space of $t$ such that $t > t'$.  Now, let us consider, $L^0 = (L^0_1, 
\ldots, L^0_n) = x^* - pe$ and $A^0 = (A_1^0, \ldots, A_n^0) = x^* + pe$. Recall that $p$ and $D$ are scalars and $e$ is a vector with all components equal to 1. From now on we will use $p$ and $D$ to denote $pe$ and $De$ respectively. Let us define two sequences ${A^k}$ and ${L^k}$ such that 

\begin{equation}\label{Eq:Ukequation}
    A^{k+1} = \frac{A^k + F(A^k) + D}{2}, \quad k \geq 0
\end{equation}

and 
\begin{equation*}
    L^{k + 1} = \frac{L^k + F(L^k) - D}{2}, \quad k \geq 0
\end{equation*}

\begin{lemm}\label{lemm:monotonicity}
For every $k \geq 0$, we have 
\begin{equation}\label{eq:uk}
    F(A^k) \leq A^{k+1} \leq A^k + D, 
 \end{equation}
 and
 \begin{equation}\label{eq:lk}
     F(L^k) \geq L^{k+1} \geq L^k - D
 \end{equation}

\end{lemm}

\begin{proof}

The proof is by induction on $k$. 

Consider, 

\begin{equation}\label{eq:U0inequality}
    \begin{array}{l}
         F(A^0) = F (x^* + p) 
         \\
         \leq F(x^*) + p 
         \\
         \leq x^* + p + D
         \\
         \leq A^0 + D
    \end{array}
\end{equation}

The second step is from Assumption \ref{assump:monotone}(d) and the third step is from Assumption \ref{assump:monotone}(c). 

Now, 

\begin{equation*}
    \begin{array}{l}
    
        A^1 = \frac{A^0 + F(A^0) + D}{2}
        \\ \\
        \leq \frac{A^0 + A^0 + D + D}{2}
        \\ \\
        = A^0 + D
        
    \end{array}
\end{equation*}
In the above equation, we have applied the inequality for $F(A^0)$ obtained in Eq.\  \ref{eq:U0inequality}.

Let us assume that 
\begin{equation*}
    A^{k+1} \leq A^k + D
\end{equation*}

This will imply, due to the monotonicity assumption, that

\begin{equation*}
    F(A^{k+1}) \leq F(A^k + D)
     \leq F(A^k) + D 
\end{equation*}

Now consider, 

\begin{equation}\label{eq:Uk2equation}
    A^{k+2} = \frac{A^{k+1} + F(A^{k+1}) + D}{2}
\end{equation}

and 

\begin{equation*}
\begin{array}{l}
    A^{k+1} = \frac{A^{k} + F(A^{k}) + D}{2}
    \\ \\
     A^{k+1} + D = \frac{A^{k} + D + F(A^{k}) + D + D}{2}
     \\ \\
     A^{k+1} + D \geq \frac{A^{k+1} + F(A^{k + 1}) + D}{2}
     \\ \\
     A^{k+1} + D \geq A^{k+2}
     \end{array}
\end{equation*}

This proves that $A^{k+1} \leq A^{k} + D$.

From the definition of $A^{k+1}$, we find that

\begin{equation*}
\begin{array}{l}
   A^{k+1} = \frac{A^{k} + F(A^{k}) + D}{2}
   \\ \\
   2A^{k+1} \geq A^{k+1} + F(A^{k})
   \\ \\
   A^{k+1} \geq F(A^{k})
\end{array}
\end{equation*}

Hence, we find that $F(A^{k}) \leq A^{k+1}$, proving Eq.\  \ref{eq:uk}. 

Using an entirely symmetrical argument, we can also prove that Eq.\  \ref{eq:lk} is true. 
\end{proof}

\begin{lemm}\label{lemm:convergence}
The sequence ${A^k}$ will converge to a point upper bounded by $x^* + 2D$ and the sequence $L^k$ will converge to a point lower bounded by $x^* - 2D$. 
\end{lemm}

\begin{proof}

We will first show that the sequence $A^k$ remains bounded from below by $x^* - D$. That is, we will show that $A^k \geq x^* - D$ for all $k$. This is true for $A^0$, by definition. Suppose that $A^k \geq x^* - D$. Then, by monotonicity and Assumption \ref{assump:monotone}(c), $F(A^k) \geq F(x^* - D) \geq x^* - D$, from which the inequality $A^{k+1} \geq F(A^k)  \geq x^* - D$ follows (from Lemma \ref{lemm:monotonicity}).

Let us consider a $\mathcal{B}$ such that, $\mathcal{B} \gg D$ and $\mathcal{B} > p$. Now, since $D$ is the bound of the $F$ mapping, but $p$ is arbitrary, we can find such a $\mathcal{B}$ for a small $D$ (we will show that $D$ is small in our application in Theorem \ref{theorem:boundtheorem}). Our objective is to prove that the sequence $A^k$ is upper bounded by $x^* + \mathcal{B}$. This is true for $A^0$ by definition. Let us assume that $A^k \leq x^* + \mathcal{B}$. Then the inequality $A^k + D \leq x^* + \mathcal{B} + D \approx x^* + \mathcal{B}$ follows. This implies that $A^{k+1} \leq A^k + \mathcal{B}$ from Lemma \ref{lemm:monotonicity}. Hence the sequence $A^k$ also has an upper bound. 

Now we have from Lemma \ref{lemm:monotonicity}, 

\begin{equation*}
    \begin{array}{l}
         A^{k+1} \leq A^k + D
         \\
         A^{k+2} \leq A^{k+1} + D
    \end{array}
\end{equation*}

Subtracting we get, 

\begin{equation*}
    A^{k+2} - A^{k+1} \leq A^{k+1} - A^{k}
\end{equation*}

This shows that $A^k$ sequence has its first difference reducing. So either this sequence should converge to a point or it should diverge to infinity (it cannot oscillate). Now, since we proved that $A^k$ is upper bounded and lower bounded by some value, it has to converge to a point. Let that point be $A^*$. 

We have from Eq.\  \ref{Eq:Ukequation}:

\begin{equation*}
    \begin{array}{l}
         A^* = \frac{A^* + F(A^*) + D}{2}
         \\ \\ 
         2A^* \leq A^* + x^* + D + D
         \\ \\
         A^* \leq x^* + 2D
    \end{array}
\end{equation*}

In the above equation, we used the fact that $F$ is upper bounded by $x^* + D$. 

Thus, the fixed point for the sequence $A^k$ is upper bounded by $x^* + 2D$. 

Using a completely symmetrical argument, we can prove that the fixed point of $L^k$ is lower bounded by $x^* - 2D$. 

\end{proof}

We will now show that for every $k$, there exists some time $t_k$ such that, 

\begin{equation}\label{eq:lkukinequality}
    L^k \leq x(t) \leq A^k, \quad \forall t \geq t_k
\end{equation}

Once this is proved, $x(t)$ can be shown to remain in the bound $x^* - 2D$ and $x^* + 2D$ from Lemma \ref{lemm:convergence}. 
We can see that Eq.\  \ref{eq:lkukinequality} is true for $k=0$, with $t_0 = 0$, by definition of $A^0$ and $L^0$. Proceeding with induction on $k$, we fix some $k$ and assume that there exists some $t_k$ so that the Eq.\  \ref{eq:lkukinequality} holds. Let $t'_k$ be such that for every $t \geq t'_k$ for every $i,j$, we have $\tau_j^i(t) \geq t_k$. Such a $t'_k$ exists because of Assumption \ref{assumption:oldinformation}.

In particular, we have 

\begin{equation*}
    L^k \leq x^i(t) \leq A^k, \quad \forall t \geq t'_k 
\end{equation*}

Let $W_i(0) = 0$ and 

\begin{equation*}
    W_i(t+1) = (1 - \alpha_i(t))W_i(t) + \alpha_i(t) w_i(t), \quad t \geq t_0
\end{equation*}

Now, we can rearrange the terms to get 

\begin{equation*}
    W_i(t+1) - \alpha_i(t) w_i(t) = (1 - \alpha_i(t))W_i(t) 
\end{equation*}

This sequence will converge to 0 in the limit ($t \xrightarrow{} \infty$) by the condition on $\alpha_i(t)$ from Assumption \ref{assumption:learningrate}. This assumption implies that 

\begin{equation*}
    \Pi_{\tau = 0}^{\infty} (1 - \alpha_i(\tau)) = 0
\end{equation*}

Thus the sequence $W_i$ will converge to $\alpha_i(t) w_i(t)$. Now, by Assumption \ref{assumption:assorted}, we can redefine $w_i(t)$ such that it goes to 0 in the limit ($t \xrightarrow{} \infty$). We then have $\lim_{t \rightarrow \infty} W_i(t) = 0$. 

For any time $t_0$, we also define $W_i(t_0; t_0) = 0$ and 

\begin{equation}\label{eq:Wequation}
    W_i(t + 1; t_0) = (1 - \alpha_i(t)) W_i(t;t_0) + \alpha_i(t) w_i(t), \quad t \geq t_0
\end{equation}

For every $t_0$ then we can have $\lim_{t \rightarrow \infty} W_i (t; t_0) = 0$ using a similar argument as above (also see Lemma 2 in Tsitsiklis \cite{tsitsiklis1994asynchronous}). 

We also define a sequence $X_i(t), t\geq t'_k$ by letting $X_i(t'_k) = A^k_i$ and 

\begin{equation}\label{eq:Xequation}
    X_i(t+1) = (1 - \alpha_i(t)) X_i(t) + \alpha_i(t) F_i(A^k), \quad  t \geq t'_k
\end{equation}

\begin{lemm}\label{lemm:Xilemma}
\begin{equation*}
x_i(t) \leq X_i(t) + W_i(t;t'_k), \forall t \geq t'_k    
\end{equation*}

\end{lemm}

\begin{proof}

Refer to Lemma 6 in Tsitsiklis \cite{tsitsiklis1994asynchronous} for proof. 
\end{proof}

We define a $\delta_k$ be equal to minimum of $(A^k_i + 2D - F_i(A^k))/4$, where the minimum is taken over all $i$ for which $(A^k_i + 2D - F_i(A^k))$ is positive. $\delta_k$ is well defined and positive due to Lemma \ref{lemm:convergence} and the lower bound of $A^k$ and upper bound of $F$. 

Let us define a $t^{''}_k \geq t'_k$, 

\begin{equation}\label{eq:alphacontraint}
    \Pi^{t^{''}_k -1}_{\tau = t'_k} (1 - \alpha_i(\tau)) \leq \frac{1}{4}
\end{equation}

and for all $t \geq t^{''}_k$ and all $i$.

\begin{equation}\label{eq:Wconstraint2}
    W_i(t;t'_k) \leq \delta_k
\end{equation}

 The condition in Eq.\  \ref{eq:alphacontraint} is possible due to Assumption \ref{assumption:learningrate}. The condition in Eq.\  \ref{eq:Wconstraint2} is possible because $W_i(t;t'_k)$ from Eq.\  \ref{eq:Wequation} converges to 0 as discussed earlier. 

\begin{lemm}\label{lemm:boundedlemma}
We have $x_i(t) \leq A_i^{k+1}$, for all $i$ and $t\geq t^{''}_k$ 
\end{lemm}

\begin{proof}
Eq.\  \ref{eq:Xequation} and the relation $X_i(t'_k) = A^k_i$ makes the process $X_i$ a convex combination of $A_i^k$ and $F_i(A^k)$. The coefficient of $A_i^k$ is equal to $\Pi_{\tau = t'_k}^{t-1} (1 - \alpha(\tau))$, whose maximum value is $\frac{1}{4}$. It follows that 

\begin{equation*}
    \begin{array}{l}
    
    X_i(t) \leq
    \\ \\ 
    \frac{1}{4} A_i^k + \frac{3}{4}F_i(A^k) = \frac{1}{2}A_i^k + \frac{1}{2} F_i(A^k) + \frac{D}{2} - \frac{1}{4}(A^k_i - F_i(A^k)) - \frac{D}{2} 
    \\ \\
     X_i(t) \leq \frac{1}{2}A_i^k + \frac{1}{2} F_i(A^k) + \frac{D}{2} - \frac{1}{4}(A^k_i - F_i(A^k) + 2D)
    \\ \\
    X_i(t) \leq A_i^{k+1} - \delta_k
    \\ \\ 
   X_i(t)  \leq A_i^{k+1} - W_i(t;t'_k)
    
    \end{array}
\end{equation*}

Now, using the result in Lemma \ref{lemm:Xilemma}, 

\begin{equation*}
    \begin{array}{l}
         x_i(t) \leq X_i(t) + W_i(t;t'_k)
         \\ \\ 
         x_i(t) \leq A_i^{k+1} - W_i(t;t'_k) + W_i(t;t'_k)
         \\ \\ 
         x_i(t) \leq A_i^{k+1}
    \end{array}
\end{equation*}

This implies that $x_i(t) \leq A_i^{k+1}$ for all $t \geq t^{''}_k$. 

\end{proof}

By an entirely symmetrical argument, we can also establish that $x_i(t) \geq L_i^{k+1}$ for all $t$ greater than some $t^{'''}_k$. This concludes the proof of the theorem.

\end{proof}

\begin{theorem2}
The MFQ mean action and the POMFQ mean action both satisfy 
\begin{equation*}
    | \Tilde{a}_{i,t} - \overline{a}_{i,t}| \leq \sqrt{\frac{1}{2n} \log \frac{2}{\delta}}
\end{equation*}
as time $t \xrightarrow{} \infty$, with probability $>= \delta$, where $n$ is the number of samples observed. $\Tilde{a}$ is the mean action as obtained from the Dirichlet in Eq.\  \ref{eq:meanaupdate} and $\overline{a}$ is the mean action in Eq.\  \ref{eq:updatemeana}. 
\end{theorem2}

\begin{proof}
Using the Hoeffding's inequality, if a set of random variables $(X_1, \cdots, X_n)$ are bounded by the intervals $[a_i, b_i]$, then the following is true: 

\begin{equation}\label{eq:hoeffding's bound}
    P(|\overline{X} - \E[\overline{X}]| \geq u) \leq 2 \exp \Big(\frac{-2n^2 u^2}{\sum_{i=1}^n (b_i - a_i)^2} \Big)
\end{equation}

\noindent
where $n$ is the number of samples.  $P$ denotes the probability and $u$ is an arbitrary bound. 

Eq.\  \ref{eq:meanaupdate} samples some $\Tilde{a}$ to estimate the partially observable Q function $Q_{POMF}$, as seen in update Eq.\  \ref{eq:valueupdate}. This is set to be $\overline{X}$ in Eq.\  \ref{eq:hoeffding's bound} and its expected value then will be true mean field action $\overline{a}$. In this setting, we have assumed that all agents have global state availability. Therefore all the agents in the environment are visible and all actions of the agents can be used to update the Dirichlet distribution, which holds the estimate of POMFQ mean action. Note that in each step, a very large number of agent actions are visible (we assume there are $N$ agents in the environment where $N$ is very large). Thus, the Dirichlet mean will become close to the true underlying $\overline{a}$. Since the agent is taking only a finite sample from the Dirichlet to update the POMF $Q$-function, the empirical mean will be $\Tilde{a}$.

Therefore, from Eq.\  \ref{eq:hoeffding's bound} we have, 

\begin{equation*}
    P(|\Tilde{a}_{i,t} - \overline{a}_{i,t}| \geq u) \leq 2 \exp \Big(\frac{-2n^2 u^2}{\sum_{j=1}^n (b_j - a_j)^2} \Big)
\end{equation*}

The samples are in the range $[0,1]$ and therefore we set $b_j = 1$ and $a_j = 0$  We set the right hand side of Eq.\  \ref{eq:hoeffding's bound} to $\delta$ to get the relation

\begin{equation*}
\begin{array}{l}
    u = \sqrt{\frac{1}{2n} \log \frac{2}{\delta}}
 \end{array}   
\end{equation*}

which proves the theorem. 
\end{proof}

\begin{theorem2}
When the $Q$-function is Lipschitz continuous (with constant M) with respect to mean actions, then the POMF $Q$-function will satisfy the following relationship: 
\begin{equation*}
    |Q^{POMF}(s_t, a_t, \Tilde{a}_{t-1}) - Q^{MF} (s_t, a_t, \overline{a}_{t-1})| \leq M \times L \times \log \frac{2}{\delta}\times \frac{1}{2n}
\end{equation*}
\noindent
as $t \xrightarrow{} \infty$ with probability $\geq (\delta)^{L-1}$, where $L=|A|$ and $n$ is the number of samples.  

\end{theorem2}

\begin{proof}

Consider a  $Q$-function that is Lipschitz continuous for all $\overline{a}$ and $\Tilde{a}$. Then we get, 

\begin{equation*}
    |Q(s_t, a_t, \Tilde{a}_{t-1}) - Q (s_t, a_t, \overline{a}_{t-1})| \leq M |\Tilde{a}_{t-1} - \overline{a}_{t-1}|
\end{equation*}

Now, from Theorem \ref{theorem:abound}, we get, 

\begin{equation*}
\begin{array}{l}
    |Q(s_t, a_t, \Tilde{a}_{t-1}) - Q (s_t, a_t, \overline{a}_{t-1})| \leq M \times L \times (\sqrt{\frac{1}{2n} \log \frac{2}{\delta}})^2
 \end{array}   
\end{equation*}

In the first step, we are taking the magnitude of the difference between the mean action vectors, hence we multiply the bound in Theorem \ref{theorem:abound} with all the components of the vectors ($\overline{a}$ and $\Tilde{a}$). The total number of components are equal to the action space $L$. Since the bound for Theorem \ref{theorem:abound} is with probability $\geq \delta$ the probability of this theorem would be at least $\delta^{L-1}$, since we have $L$ random variables, and when we fix the first $L-1$ random variables, the last one is deterministic as all the components of $\overline{a}$ satisfy the relation that sum of the individual components will be 1 (one hot encoding). All the components of $\Tilde{a}$ also satisfy this relation as $\Tilde{a}$ is a normalized sample obtained from the Dirichlet.  

Then, from the definition of POMF $Q$-function and MFQ $Q$-function, the theorem follows.

\end{proof}

\begin{theorem2}
When we update the $Q$ functions using the partially observable update rule in Eq.\  \ref{eq:qupdate}, the process
satisfies the condition in the limit ($t \rightarrow \infty$):
\begin{equation*}
    |Q^*(s_t, \boldsymbol{a}_t) - Q^{POMF}(s_t, a_t, \Tilde{a}_t)| \leq 2D
\end{equation*}
\noindent
when Assumptions \ref{assumption:learningrate}, \ref{assumption:rewardbound}, and \ref{assumption:globaloptimum} hold. Here $Q^*$ is the Nash Q-value and $D$ is the bound for value functions in Eq.\  \ref{eq:valuebound}. This holds with probability at least $\delta^{L-1}$, where $L=|A|$.      

\end{theorem2}

\begin{proof}
We start by proving all the assumptions needed for Theorem \ref{theorem:tsitskilis2} and then apply Theorem \ref{theorem:tsitskilis2} to prove this theorem.

We can write the $Q$ update from Eq.\  \ref{eq:qupdate} using the formula: 

\begin{equation}\label{eq:newupdate}
\begin{array}{l}
    Q^{POMF}_{t+1}(s_t^j, a_t^j, \Tilde{a}_t^j) = \\ \\ Q^{POMF}_{t}(s_{t}^j, a_{t}^j, \Tilde{a}_t^j) + \alpha_{t}[r^j_t + \gamma v^{POMF}_{t}(s^j_{t+1}) - Q^{POMF}_{t}(s_t^j, a_t^j, \Tilde{a}_t^j)] 
    \end{array}
\end{equation}

Let $F$ be defined as

\begin{equation}\label{eq:Fupdate}
    F(Q^{POMF}_{t}(s_{t+1}^j, a_t^j, \Tilde{a}_t^j)) = \E[r^j] + \gamma \E[v^{POMF}_{t}(s_{t+1}^j)]
\end{equation}

and the value function from Eq.\  \ref{eq:valueupdate} be

\begin{equation}\label{eq:newvalueupdate}
    v^{POMF}_{t}(s^j_{t+1}) = \sum_{a^j}\pi^j_t(a^j|s',\Tilde{a}^j)  Q_t^{POMF}(s^j_{t+1},a^j, \Tilde{a}^j)
\end{equation}

Using Eq.\  \ref{eq:Fupdate}, then Eq.\  \ref{eq:newupdate} can be written as 

\begin{equation*}
\begin{array}{l}
    Q^{POMF}_{t+1}(s^j_t, a^j_t, \Tilde{a}^j_t) = Q^{POMF}_{t}(s^j_t, a^j_t, \Tilde{a}^j_t) + \\ \\
    \alpha_{t}[F(Q^{POMF}_{t}(s^j_t, a^j_t, \Tilde{a}^j_t)) - Q^{POMF}_{t}(s^j_t, a^j_t, \Tilde{a}^j_t) + w_{t}(s^j_t, a^j_t, \Tilde{}{a}^j_t)]
    \end{array}
\end{equation*}

\noindent 
where 

\begin{equation*}
    w_{t}(s^j_t, a^j_t, \Tilde{a}_t^j) = r^j - \E[r^j] + \gamma v^{POMF}_{t}(s^j_{t+1}) -  \gamma \E[v^{POMF}_{t}(s^j_{t+1})]
\end{equation*}

Assumption \ref{assumption:oldinformation} is satisfied, since we are setting $\tau^i_j(t) = t$ in Eq.\  \ref{eq:newupdate}. Our formulation is the same as that in Tsitsiklis \cite{tsitsiklis1994asynchronous} where the $Q$ function is not allowed any outdated information from the previous steps in the current update. Assumption \ref{assumption:oldinformation} only needed to guarantee that, if old information were used, that would be eventually discarded w.p.1. 

To satisfy all the measurability conditions of Assumption 2, let $\mathcal{F}_t$ be a $\sigma$-field generated by random variables in all history of the stochastic game till $t$: $(s_t, \alpha_t, \boldsymbol{a}_t, r_{t-1}, \tau_t, \cdots, s_1, \alpha_1, \boldsymbol{a}_1, \tau_1, Q_0)$. Let $Q_t$ be a random variable from this trajectory and is hence $\mathcal{F}_t$-measurable. Thus, from Eq.\  \ref{eq:Fupdate}, we can see that $F$ will also be $\mathcal{F}_t$-measurable. This satisfies Assumptions \ref{assumption:assorted}(a), (b), and (c). Assumption \ref{assumption:assorted}(d) can be directly verified from Eq.\  \ref{eq:Fupdate}. From Assumption \ref{assumption:rewardbound} it can be shown that the action-value function and the value function remain bounded and hence we can find arbitrary constants $A$ and $B$ such that Assumption \ref{assumption:assorted}(e) will hold.

To prove Assumption \ref{assump:monotone}(a), note that the
$F$ mapping in Eq.\  \ref{eq:Fupdate} will increase or decrease only as the term $v^{POMF}$ increases or decreases for a given reward function. Now $v^{POMF}$ depends on $Q^{POMF}$.  As $Q^{POMF}$ increases or decreases then $v^{POMF}$ also increases or decreases for a stationary policy. Thus the monotonicity condition (Assumption \ref{assump:monotone}(a)) is proved. Because the function is a linear function, the continuity condition is also proved (Assumption \ref{assump:monotone}(b)).
Hence, the first two conditions of Assumption \ref{assump:monotone} are satisfied.

Using Eq.\  \ref{eq:Fupdate} we can write: 

\begin{equation*}
\begin{array}{l}
    F(Q^{POMF}_{t}(s_t^j, a_t^j, \Tilde{a}_t^j)) 
    = \\ \\
    \E[r^j] + \gamma \E[v^{MF}_{t}(s_{t+1}^j)] + \gamma [\E[v^{POMF}_{t}(s_{t+1}^j)] - \E[v^{MF}_{t}(s_{t+1}^j)]]
    \\ \\
    = \E[r^j + \gamma [v^{Nash}_{t}(s_{t+1}^j)]] + 
    \E (\gamma[v^{POMF}_{t}(s_{t+1}^j) - v^{MF}_{t}(s_{t+1}^j)])
    \\ \\
    + \E (\gamma[v^{MF}_{t}(s_{t+1}^j) - v^{Nash}_{t}(s_{t+1}^j)])
    \leq 
     Q^*(s^j_t,\boldsymbol{a}^j_t) + D
\end{array}
\end{equation*}

Also we have, 
\begin{equation*}
\begin{array}{l}
    F(Q^{POMF}_{t}(s_t^j, a_t^j, \Tilde{a}_t^j)) 
    \\ \\
    = \E[r^j + \gamma [v^{Nash}_{t}(s_{t+1}^j)]] + \E (\gamma[v^{POMF}_{t}(s_{t+1}^j) - v^{MF}_{t}(s_{t+1}^j)]) - 
    \\ \\
    \E[\gamma [v^{Nash}_{t}(s_{t+1}^j) - v^{MF}_{t}(s_{t+1}^j) ]]
    \geq 
    Q^*(s^j_t,\boldsymbol{a}^j_t) - D
\end{array}
\end{equation*}

In the above two equations we use the fact that in the limit ($t \xrightarrow{} \infty$) the mean field value function becomes equal to the Nash value function given Assumption \ref{assumption:globaloptimum} (see Theorem 1 in Yang et al.\ \cite{pmlr-v80-yang18d} for the proof). Thus, that term can be dropped. This proves Assumption \ref{assump:monotone}(c).

Now to prove Assumption \ref{assump:monotone}(d), consider,

\begin{equation}\label{eq:assumption4update0}
\begin{array}{l}
    F(Q^{POMF}_{t}(s_t^j, a_t^j, \Tilde{a}_t^j) + p) 
    \\ \\
    = \E[r^j] + \gamma [X  Q_t^{POMF}(s^j_{t+1},a^j, \Tilde{a}^j) + p] 
    \\ \\
    = \E[r^j] + \gamma [X  Q_t^{POMF}(s^j_{t+1},a^j, \Tilde{a}^j)] + \gamma p] 
\end{array}
\end{equation}

where $X = \sum_{a^j} \pi^j_t (a^j|s', \overline{a}^j)$. \\

Now consider, 

\begin{equation}\label{eq:assumption4update1}
    \begin{array}{l}
         F(Q^{POMF}_{t}(s_t^j, a_t^j, \Tilde{a}_t^j)) + p
         \\ \\
         = \E[r^j] + \gamma [X Q_t^{POMF}(s^j_{t+1},a^j, \Tilde{a}^j)] + p
    \end{array} 
\end{equation}
Since, $\gamma \leq 1$, from Eqs.\  \ref{eq:assumption4update0} and \ref{eq:assumption4update1}, we find that

\begin{equation}\label{eq:assumption4update3}
   F(Q^{POMF}_{t}(s_t^j, a_t^j, \Tilde{a}_t^j) + p) \leq F(Q^{POMF}_{t}(s_t^j, a_t^j, \Tilde{a}_t^j)) + p
\end{equation}

By a symmetric argument, we can prove  $$F(Q^{POMF}_{t}(s_t^j, a_t^j, \Tilde{a}_t^j)) - p \leq F(Q^{POMF}_{t}(s_t^j, a_t^j, \Tilde{a}_t^j) - p)$$ and the condition  $$ F(Q^{POMF}_{t}(s_t^j, a_t^j, \Tilde{a}_t^j) - p) \leq F(Q^{POMF}_{t}(s_t^j, a_t^j, \Tilde{a}_t^j) + p) $$ 

will hold by monotonicity.

Thus, all the conditions of Assumption \ref{assump:monotone} have been proved. 

Consider the POMF $Q$-update in Eq.\  \ref{eq:qupdate}. Notice that if the reward function $r$ is guaranteed to be bounded, then the $Q$-function will also be bounded as there are no other variables that $Q$ depends on, which can make it diverge to infinity. 
Since Assumption 5 guarantees that the reward function will stay bounded, the value and action-value functions will also stay bounded. This will mean that the entire stochastic process as given in Theorem \ref{theorem:tsitskilis2} remains bounded. 
Now, with all conditions met, we can prove that the partially observable mean field $Q$ function will either converge or oscillate in a small range $(Q^*(s_t, \boldsymbol{a}_t) - 2D, Q^*(s_t, \boldsymbol{a}_t) + 2D)$ in the limit ($t \xrightarrow{} \infty$) according to Theorem \ref{theorem:tsitskilis2}. 
\end{proof}

\section{Recurrent baselines}\label{sec:recurrentbaselinesappendix}

This section describes our comparisons of the POMFQ - FOR and PDO algorithms with recurrent versions of IL and MFQ (denoted as RIL and RMFQ respectively). We report all results for 20 independent runs as done for the comparisons in Section \ref{sec:expertiments}. 

\begin{figure}
	\subfloat[FOR -  Train]{{\includegraphics[width=0.22\textwidth, height=4cm]{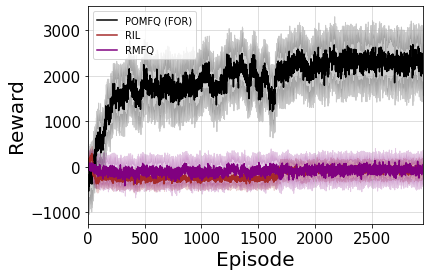} }}
	\subfloat[PDO -  Train]{{\includegraphics[width=0.22\textwidth, height=4cm]{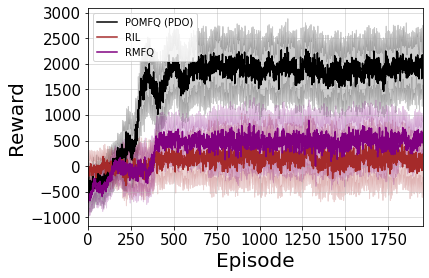} }}
	\\
	\subfloat[FOR -  Test]{{\includegraphics[width=0.22\textwidth, height=3.8cm]{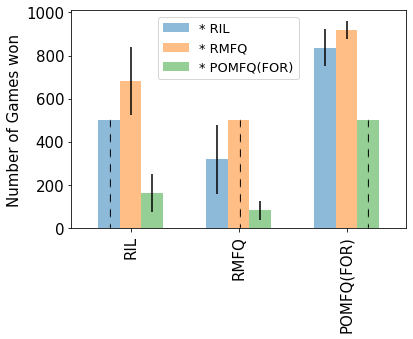} }}
	\subfloat[PDO -  Test]{{\includegraphics[width=0.22\textwidth, height=3.8cm]{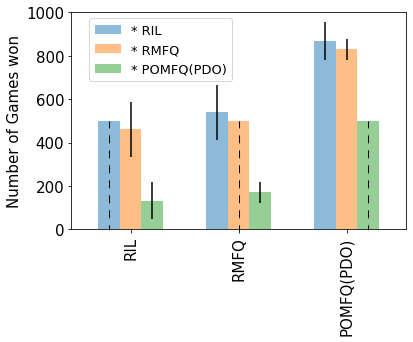} }}
  \caption{Multibattle results with recurrent baselines}%
	\label{fig:multibattlerecurrent}
\end{figure}

Multibattle game results are given in Figure \ref{fig:multibattlerecurrent}. In the FOR domain, Figure~\ref{fig:multibattlerecurrent}(a), we can see that the recurrent algorithms fall into a local optimum early on and do not improve much, similar to the performance of MFAC. The reason for this could be that the recurrent network provides a larger amount of information to the agent at each time step because the state is recurrently fed back to the network as an input. In a mean field setting, this may not a good idea since there is already an overload of information available to each agent, and the superior performance of POMFQ is due to the fact that it supplies exactly the right amount of information needed to calculate the best response at each time step. In the test battles for FOR (Figure \ref{fig:multibattlerecurrent} (c)), POMFQ (FOR) comfortably beats RIL (p < 0.01) and RMFQ (p < 0.01). This performance advantage is also seen in the PDO setting, where the POMFQ (PDO) beats RIL and RMFQ in both the train (Figure \ref{fig:multibattlerecurrent}(b), p < 0.01) and test experiments (Figure \ref{fig:multibattlerecurrent}(d), p < 0.01).

\begin{figure}
	\subfloat[FOR -  Train]{{\includegraphics[width=0.22\textwidth, height=4cm]{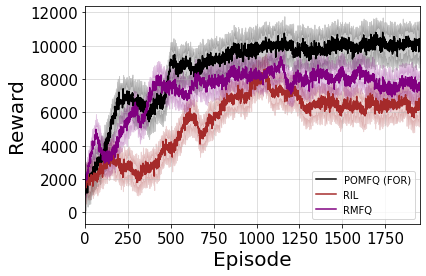} }}
	\subfloat[PDO -  Train]{{\includegraphics[width=0.22\textwidth, height=4cm]{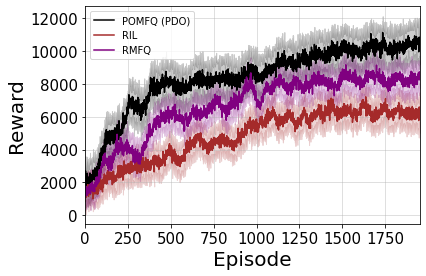} }}
	\\
	\subfloat[FOR -  Test]{{\includegraphics[width=0.22\textwidth, height=3.8cm]{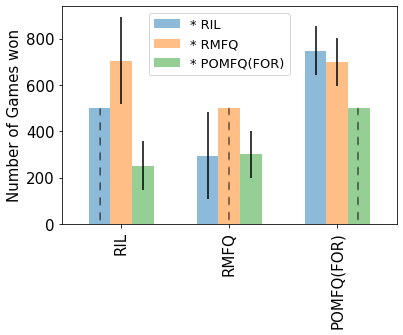} }}
	\subfloat[PDO -  Test]{{\includegraphics[width=0.22\textwidth, height=3.8cm]{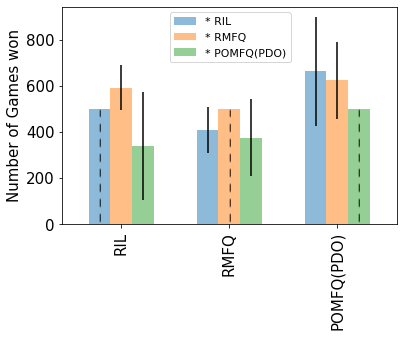} }}
  \caption{Battle-Gathering results with recurrent baselines}%
	\label{fig:multigatherrecurrent}
\end{figure}

In the Battle-Gathering (Figure \ref{fig:multigatherrecurrent}) we still see that the POMFQ beats the recurrent baselines in both training and test performances in both the domains. However, it is interesting to note that in both the FOR (Figure \ref{fig:multigatherrecurrent} (a)) and PDO (Figure \ref{fig:multigatherrecurrent} (b)) experiments, both the recurrent baselines learn to perform reasonably well and do not fall into a trivial local optimum like in the Multibattle game. Recall that this game has two tasks, one is to gather as many food particles as possible and the other is to kill the enemies. The recurrent baselines seem to learn to perform one of the tasks very well (gathering food), but still lose out to POMFQ, which learns to perform both the tasks reasonably well. The test results in Figure \ref{fig:multigatherrecurrent}(c) and (d) also show the performance advantage of POMFQ against the recurrent baselines in this domain.  The train results have p < 0.04 and test results have p < 0.02 for the comparisons between POMFQ and the next best performing recurrent algorithm. 

\begin{figure}
	\subfloat[FOR -  Train]{{\includegraphics[width=0.22\textwidth, height=4cm]{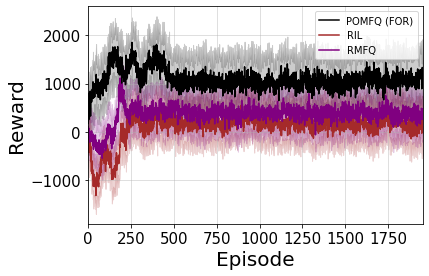} }}
	\subfloat[PDO -  Train]{{\includegraphics[width=0.22\textwidth, height=4cm]{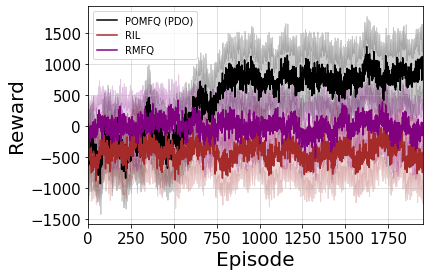} }}
	\\
	\subfloat[FOR -  Test]{{\includegraphics[width=0.22\textwidth, height=3.8cm]{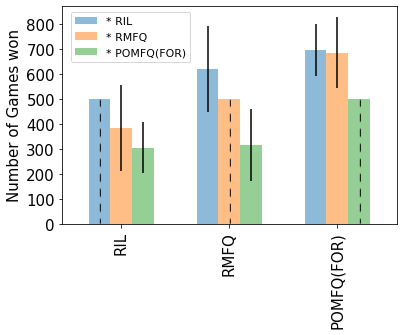} }}
	\subfloat[PDO -  Test]{{\includegraphics[width=0.22\textwidth, height=3.8cm]{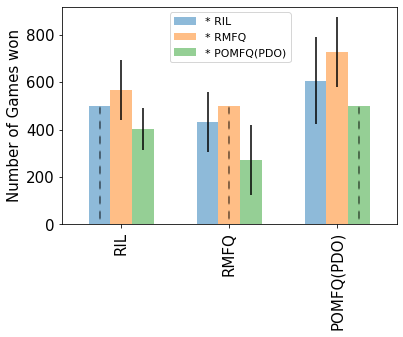} }}
  \caption{Predator-Prey results with recurrent baselines}%
	\label{fig:predatorpreyrecurrent}
\end{figure}

For the Predator-Prey domain, the results are given in Figure \ref{fig:predatorpreyrecurrent}. In this domain too, POMFQ algorithm beats the performance of both the recurrent baselines in the training experiments (Figure \ref{fig:predatorpreyrecurrent}(a) and (b) with p < 0.06) as well as the test experiments (Figure \ref{fig:predatorpreyrecurrent}(c) and (d)) with p < 0.03). Like comparisons with the other algorithms for the Predator-Prey domain, our training results are not statistically significant but the test results are statistically significant.

\section{Ising model}\label{sec:isingappendix}

\begin{figure}
    \centering
    \includegraphics[width=0.4\textwidth ]{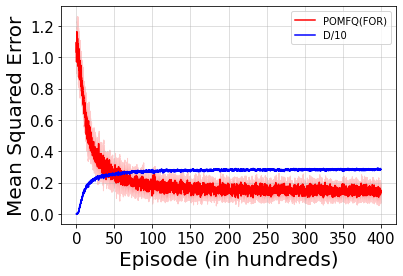}
    \caption{Ising model: Mean square distance between the $Q$-values and Nash $Q$-values}
    \label{fig:ising}
\end{figure}

The Ising model was introduced as a stochastic game by Yang et al. \cite{pmlr-v80-yang18d}. The energy function of this model determines that the overall energy of the system stays low when each agent chooses to spin in a direction that is consistent with its neighbours. In this game, each agent has a choice of one of the two actions (spinning up or spinning down) and obtains rewards proportional to the number of agents spinning in the same direction in the neighbourhood. We use the same settings and parameters for the Ising model as in Yang et al. \cite{pmlr-v80-yang18d}. Particularly, each agent obtains rewards [-2.0, -1.0, 0.0, 1.0, 2.0] based on the number of neighbours [0,1,2,3,4] spinning in the same direction as itself in each stage game. Refer to \cite{pmlr-v80-yang18d} for more details on this domain. We set the temperature $\tau$ of the ising model to 0.8.  We use this domain to give a practical illustration of the tabular version of our algorithm on a simple domain and to verify Theorem \ref{theorem:boundtheorem}. The Ising model used in our implementation is a stateless system with a total of 100 agents and the Nash $Q$-function of this stochastic game is exactly obtainable \cite{pmlr-v80-yang18d}. We implement the POMFQ (FOR) algorithm on this domain and calculate the mean square error (MSE) between the $Q$-values of each agent's action with the Nash $Q$-value at each stage. The average of this error across all the 100 agents is plotted against number of episodes in Figure~\ref{fig:ising}, along with the 95\% confidence interval shaded out around each point. From the plot it can be seen that the MSE steadily reduces and its 95\% confidence interval stays bounded by the line representing the value of $D/10$ line after a finite number of episodes. This is a stronger result than Theorem \ref{theorem:boundtheorem} where we had proved that the error will be bounded by $2D$. This shows that there is some scope for a stronger version of Theorem \ref{theorem:boundtheorem} as well. Substituting the value of $\delta$ as 0.95 in Theorem \ref{theorem:abound} we can calculate the value of the constant $D$ used in Theorem \ref{theorem:boundtheorem}. Notice that $D$ as given in Eq.\  \ref{eq:valuebound} is the sum of two other constants $Z$ and $K$. We set the value of the number of samples $n$ drawn from the Dirichlet at each time step to be 10000 and hence from Theorem \ref{theorem:POMFMFbound} it can be seen that constant $Z \approx 0$. Now to calculate the Lipschitz constant $K$, we calculate the difference between the MFQ $Q$-values between the two actions for the given state and mean action as in Eq \ref{eq:boundonQwithsamea}. Since we start all the $Q$-values with an initialization of 0, the value of $K$ (and hence, $D$) is 0 at the beginning of Figure \ref{fig:ising}. However, once learning takes place and the $Q$-values change, we see a finite value of $D$. As the $Q$-values move closer to convergence, the value of $D$ also converges to a finite value as seen in Figure \ref{fig:ising}. This value is used to compare the mean squared error of the $Q$-values in Figure \ref{fig:ising}.

\section{Experimental Details}\label{sec:experimentaldetailsappendix}

In this section, we provide more details regarding the game domains and hyperparameters of our algorithms. 

\subsection{MAgent Games}

The state space in our MAgent games do not have layers containing spatial information, as in the default state space in MAgents. We modify the state space to contain complete information (position, health, group information) about the viewable agents along with the individual agent features (already available in MAgent). All the agents at a distance of the view range from the central agent are visible in the FOR setting and for the PDO domains the viewable agents are sampled from the Bernoulli distribution, in all the games. Additionally, for the Battle-Gathering game, the position of all the food in the environment is available to all agents in their observation of the state. For all the games, we assume that at any given time step any agent cannot process more than 20 other agents and hence at most the 20 closest agents are considered. The action space for all the games includes move and attack actions. There are a total of 21 actions, with 13 move actions (move to one of the 13 cells in a unit circle, refer to \cite{zheng2018magent} for more details) and 8 attack actions (attack one of the 8 nearest cells in a unit circle).

Note that training is decentralized at the group level. Agents are independent in terms of the information that they act on: if an agent were to try to account for the group's full observation space, they would quickly become overwhelmed in this setting. Agents do not use separate neural networks for each agent as the complexity would be linear in the number of agents. Instead, agents in a group train on, and use, a single neural network. This is consistent with Yang et al.'s battle games \cite{pmlr-v80-yang18d} and the MAgent baselines \cite{zheng2018magent}.

In the faceoff experiments for the Multibattle and the Battle-Gathering games, for every 1000 game set, group A from the first algorithm and group B trained using the second algorithm fight against each other for the first 500 games and group B from the first algorithm and group A from the second algorithm fights for the second 500 games. The team that has the highest number of agents alive at the end of a game (500 steps) wins the battle. If both teams have the same number of agents alive, then the team that gets the highest reward at the end is the winner. For the Predator-Prey game, the faceoff contest is conducted similar to the Multibattle game, but with a small change in how the winners are determined. The winner is the group with the most agents alive at the end of the game. If two groups have the same agents alive, then the game is considered to be a draw. Since we start with more prey (40) than predators (20), we have a fair contest in the faceoff as the predators have to kill many more prey to win the game and the prey have to attempt to escape the predators.

The reward function for the Multibattle domain gives every agent a -0.005 reward for each step and a -0.1 for attacking an empty grid (a needless attack). The agents get a +200 for killing opponents and a +0.2 for successfully attacking. Each agent has a health of 10 which must be exhausted by damage before the agent dies. All the agents are of size 1 unit width and 1 unit height and have a speed of 2 units per turn. In the Battle-Gathering domain, agents get a +80 for capturing food and a +5 for killing opponents with all other rewards being similar to the Multibattle domain. For both the Multibattle and Battle-Gathering domains, the view range is 6 units. In the Predator-Prey domain, the predators and prey have different reward functions. The predators have sizes of width 2 units and height 2 units with a maximum health point of 10 units. They have a speed of 2 units. The prey have sizes of width 1 unit and height 1 unit and a maximum health point of 2 units. The speed of prey is 2.5 units. The speed determines the size of the circle where the cells containing a valid move direction lies (refer to \cite{zheng2018magent} for more details). The view range of predators are 7 units and the view range of prey are 6 units. The predators get a +1 for attacking prey and the prey get a -1 for being attacked. The predators receive a reward of +100 for killing prey. The predators get a -0.3 for making a needless attack. The prey get a -0.5 for dying (dead penalty). The reward function for all the three domains are same for the FOR and PDO settings. The rewards in all the MAgent games is per agent. That is, each agent gets an individual reward based on its actions in the environment. However, we sum all the rewards obtained by the agents in a team for our training plots in the paper. 

Across our three domains there is more useful information available in the local observation, which makes it easier for algorithms that do not model partial information. Unlike Multibattle, Battle-Gathering has food available in the local observation that agents can capture. Predator-Prey has more agents (60) with distinct roles (predators must only attack and prey must escape) that makes modelling partial information less important. Consequently, as seen in the experimental results, the performance gain in using POMFQ instead of MFQ is maximized in Multibattle, and decreases for Battle-Gathering and Predator-Prey. Yet, POMFQ beats MFQ in all games. This can be observed across the two settings (FOR and PDO) and in both the train and test experiments. IL loses out in Battle-Gathering compared to POMFQ as the independent strategy finds it difficult to balance the twin goals (capturing food and killing opponents).

\subsection{Hyperparameters}

The hyperparameters of IL, RIL, MFQ, RMFQ and POMFQ are almost the same. The learning rate is $\alpha = 10^{-4}$ and the exploration rate $\beta$ decays from 1 to 0 linearly during the 2000 (or 3000 as the case may be) rounds of training. The discount factor $\gamma$ is 0.95, the size of replay buffer is $2^{10}$, and the mini batch size is 64. POMFQ always takes 100 samples for all sampling steps in both the algorithms. The recurrent baselines (RIL and RMFQ) contains a GRU (gated recurrent unit) layer in addition to the fully connected layers. We take 100 samples in all sampling steps (i.e. from Dirichlet and Gamma distributions). 

MFAC has the same learning rate and batch size as the other three algorithms, the temperature of soft-max layer of actor is $\tau = 0.1$, the coefficient of entropy in the total loss is 0.08, and the coefficient of value in the total loss is 0.1. 

Most hyperparameters are same as those maintained by Yang et al.\ \cite{pmlr-v80-yang18d} in their battle game experiments. 

Each training of 2000 rounds takes a wall clock time of 18 -- 24 hours to complete on a virtual machine with 2 GPUs and 50 GB of memory. The test experiments take approximately 12 hours to complete 1000 rounds on a similar virtual machine.

\section{Ablation study}\label{sec:ablationstudyappendix}

In this section, we perform an ablation study by varying the neighbourhood distance and study the performance of the POMFQ (FOR) algorithm.

\begin{figure}[ht]
    \centering
    \includegraphics[width=0.4\textwidth ]{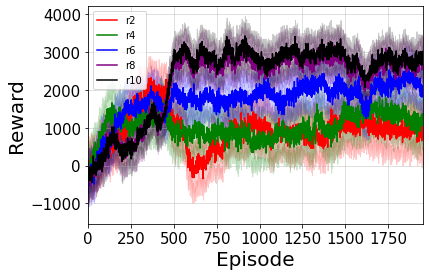}
    \caption{Multibattle (FOR case) with varying viewing distance}
    \label{fig:ablation}
\end{figure}

Figure \ref{fig:ablation} shows the results of an ablation study where we change the distance of neighbourhood view in the Multibattle game. We study the neighbourhood viewable distances of 2 units, 4 units, 6 units, 8 units, and 10 units. These are denoted as r2, r4, r6, r8, and r10 respectively in Figure \ref{fig:ablation}. Our results are averaged over 20 independent trials similar to the other experiments. The experiment is here is similar to our train experiments, where the POMFQ(FOR) algorithm competes against itself. In r2, groups A and B can see a distance of 2, and in r10, groups A and B can all see a distance of 10. When the average number of agents seen is higher (because the observation distance is higher), we expect the reward will be greater than or equal to the case where the average number of agents seen is lower due to the observation distance being lower. The Figure \ref{fig:ablation} shows that performance does steadily increase as we increase the neighbourhood distance. When more agents in the environment are seen by the central agent, more useful information is available about the environment --- as expected, this impacts performance positively.  As an example, once the agents are able to see a larger distance, they do need to indiscriminately employ the attack action that entails a penalty for usage against an empty neighbour (reward function in Appendix \ref{sec:experimentaldetailsappendix}). 

It is important to note that these games are not zero sum games (reward functions are in Appendix \ref{sec:experimentaldetailsappendix}). We have empirically seen that when agents have access to limited information they quickly fall into sub-optimal performance based on the limited neighbourhood they can see. In this case, they are calculating their actions based on limited information and  strategies are sub-optimal. When more information is available to the agents, their performance increases as they take into account a larger context before deciding best response strategies. In Figure \ref{fig:ablation}, we see that there is almost no difference in performance between r8 and r10 --- when the distance increases, the actions of agents at a large distance does not have a considerable impact on the performance of the central agent. The performance of the agent when the viewing distance is 2 units and 4 units are also not much different but overall, but the setting with distance 4 units just outperforms the setting with distance of 2 units (p < 0.4). The performance with distance 6 units is in between the performance with 4 units distance (p < 0.2) and 8 units distance (p < 0.2). Despite seeing the described performances, it should be noted that these differences are not statistically significant as p < 0.2. The difference between r2 and r10 has p < 0.01, which is statistically significant. Additionally, we see some unusual behaviour of unlearning in the case of r2. The agents could have started off being aggressive due to higher exploration but switched to more defensive strategies later on due to the availability of very less information. Nonetheless, this does not change the core message of this section.

\section{Complexity analysis}\label{sec:complexityanalysisappendix}

A tabular version of our algorithm is linear in number of states, polynomial in the number of actions, and constant in the number of agents. The guarantees are similar to the paper from Hu and Wellman \cite{hu2003nash}, except that their algorithm is exponential in the number of agents. The time complexity is also same as the space complexity as in the worst case, each entry in the table has to be accessed and updated.  

Note that the approach by Yang et al.\ \cite{pmlr-v80-yang18d} has exponential space complexity in the number of agents, since each agent has to maintain the Q tables for every other agent to obtain the action of other agents in Eq.\  \ref{eq:yangvalueupdate}. This is much worse than our space complexity.

%%%%%%%%%%%%%%%%%%%%%%%%%%%%%%%%%%%%%%%%%%%%%%%%%%%%%%%%%%%%%%%%%%%%%%%%

%%% The next two lines define, first, the bibliography style to be 
%%% applied, and, second, the bibliography file to be used.

%%%%%%%%%%%%%%%%%%%%%%%%%%%%%%%%%%%%%%%%%%%%%%%%%%%%%%%%%%%%%%%%%%%%%%%%

\end{document}